# A SURVEY ON TIDAL ANALYSIS AND FORECASTING METHODS FOR TSUNAMI DETECTION

**Sergio Consoli**[(1),*]

European Commission Joint Research Centre, Institute for the Protection and Security of the Citizen,
Via Enrico Fermi 2749, TP 680, 21027 Ispra (VA), Italy

**Diego Reforgiato Recupero**

National Research Council (CNR), Institute of Cognitive Sciences and Technologies,
Via Gaifami 18 - 95028 Catania, Italy

R2M Solution, Via Monte S. Agata 16, 95100 Catania, Italy.

**Vanni Zavarella**

European Commission Joint Research Centre, Institute for the Protection and Security of the Citizen,
Via Enrico Fermi 2749, TP 680, 21027 Ispra (VA), Italy

## ABSTRACT

Accurate analysis and forecasting of tidal level are very important tasks for human activities in oceanic and coastal areas. They can be crucial in catastrophic situations like occurrences of Tsunamis in order to provide a rapid alerting to the human population involved and to save lives. Conventional tidal forecasting methods are based on harmonic analysis using the least squares method to determine harmonic parameters. However, a large number of parameters and long-term measured data are required for precise tidal level predictions with harmonic analysis. Furthermore, traditional harmonic methods rely on models based on the analysis of astronomical components and they can be inadequate when the contribution of non-astronomical components, such as the weather, is significant. Other alternative approaches have been developed in the literature in order to deal with these situations and provide predictions with the desired accuracy, with respect also to the length of the available tidal record. These methods include standard high or band pass filtering techniques, although the relatively deterministic character and large amplitude of tidal signals make special techniques, like artificial neural networks and wavelets transform analysis methods, more effective. This paper is intended to provide the communities of both researchers and practitioners with a broadly applicable, up to date coverage of tidal analysis and forecasting methodologies that have proven to be successful in a variety of circumstances, and that hold particular promise for success in the future. Classical and novel methods are reviewed in a systematic and consistent way, outlining their main concepts and components, similarities and differences, advantages and disadvantages.

[(1)] Current affiliation: National Research Council (CNR), Institute of Cognitive Sciences and Technologies,
Via Gaifami 18 - 95028 Catania, Italy
[*] Corresponding author - Email: sergio.consoli@istc.cnr.it; Phone: (+39) 095 7338 390; Fax:(+39) 095 7338 390

# Table of Contents

# 1  Introduction

*Tides* are the rise and fall of sea levels caused by the combined effects of the gravitational forces exerted by the Moon and the Sun and the rotation of the Earth. Most places in the ocean usually experience two high tides and two low tides each day (semi-diurnal tide), but some locations experience only one high and one low tide each day (diurnal tide). The times and amplitude of the tides at the coast are influenced by the alignment of the Sun and Moon, by the pattern of tides in the deep ocean and by the shape of the coastline and near-shore bathymetry.

Figure 1 shows the range variations for the different types of tides.

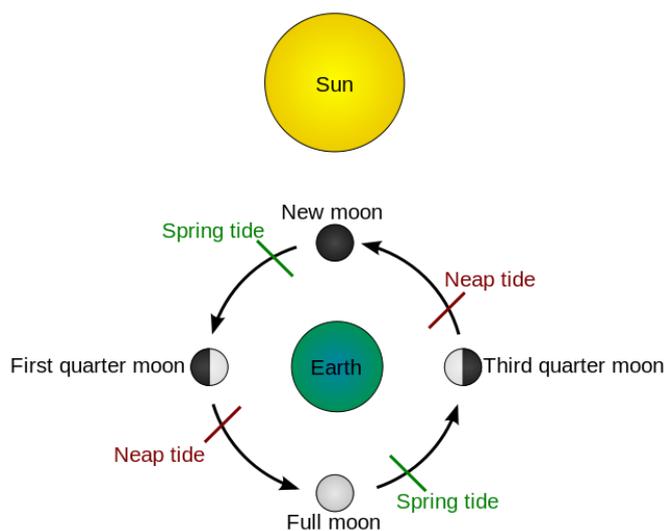

**Figure 1 - Due to the bathymetry of some areas, neap and spring tides reach their maximum force 2 days after the first quarter moon, third quarter moon and new moon, full moon, respectively (Wikipedia, 2013).**

Tides are influenced by different factors. To make accurate records, tide gauges at fixed stations measure the water level over time. Gauges ignore variations caused by waves with periods shorter than minutes. These data are compared to the reference (or datum) level usually called mean sea level. While tides are usually the largest source of short-term sea-level fluctuations, sea levels are also subject to forces such as wind and barometric pressure changes, resulting in storm surges, and to displacements of large volumes of a body of water caused by earthquakes, volcanic eruptions or other underwater explosions, generating *tsunamis* (Shiki et al., 2008).

Tsunami waves do not resemble normal sea waves, because their wavelength is far longer. Rather than appearing as a breaking wave, a tsunami may instead initially resemble a rapidly rising tide, and for this reason they are often referred to as tidal waves (although tsunamis actually have nothing to do with tides). Tsunamis generally consist of a series of waves with periods ranging from minutes to hours, arriving in a so-called "wave train" (Fradin and Brindell, 2008). Wave heights of tens of meters can be generated by large events. Although the impact of tsunamis is limited to coastal areas, their destructive power can be enormous and they can affect entire ocean basins; the 2004 Indian Ocean tsunami was among the deadliest natural disasters in human history with over 230.000 people killed in 14 countries bordering the Indian Ocean (Lay et al., 2005).

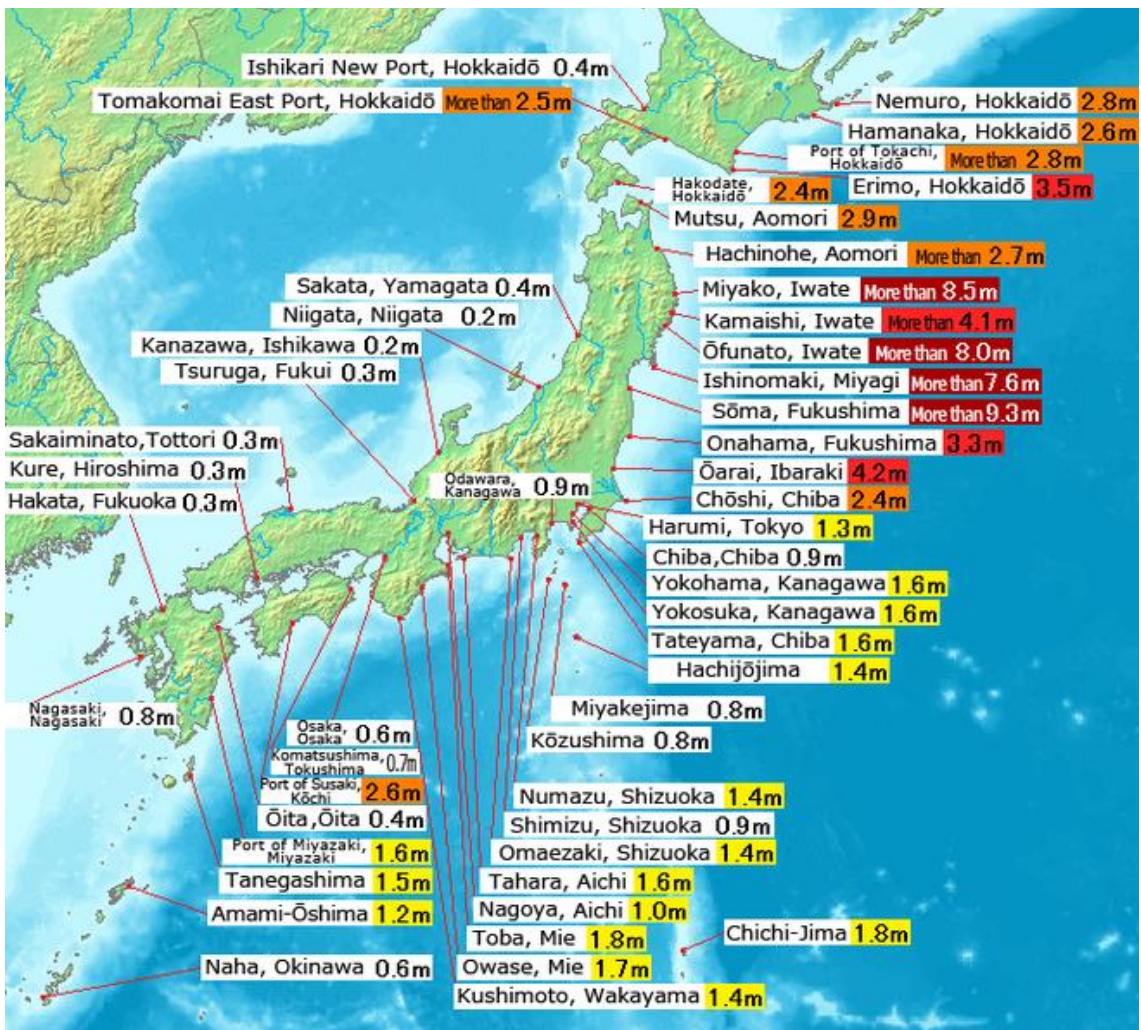
**Figure 2 - Tohoku earthquake observed tsunami heights (Wikipedia, 2013a).**

After this terrible event, other notable tsunamis have occurred. In 2006, a 7.7 magnitude earthquake rocked the Indian Ocean seabed and triggered tsunamis whose height varied from 2 to 6 meters at South of Java Island, where it swept away and flattened buildings as far as 400 meters away from the coastline, and it caused more than 800 people missed or dead (WHO, 2006).

On April 2007, a powerful magnitude 7.6 earthquake hit the East Pacific region about 40 km the western Solomon Islands, resulting in a tsunami that was up to 12 m tall (Furlong et al., 2009). The wave triggered region-wide tsunami warnings and watches extending from Japan to New Zealand to Hawaii and the eastern seaboard of Australia. The tsunami killed 52 people, and dozens more were injured, with entire towns inundated by the sweeping water. Officials estimate that the tsunami displaced more than 5000 residents all over the archipelago.

The 2009 Samoa earthquake was an 8.1 magnitude submarine earthquake that generated a tsunami with 76 mm rise in sea levels near the epicentre, and up to 14 metres waves on the Samoan coast (Atayman, 2009). More than 189 people were killed, especially children.

The 2010 Chile earthquake occurred off the coast of central Chile with a magnitude of 8.8 and an intense shaking lasting for about three minutes (UNAVCO, 2010). It ranks as the sixth largest earthquake ever to be recorded by a seismograph. The earthquake triggered a tsunami that devastated several coastal towns in south-central Chile and damaged the port at Talcahuano. The "state of catastrophe" was declared and military troops were sent to take control of the most affected areas. According to official sources, 525 people lost their lives, 25 people went missing, about 9% of the population lost their homes, and damage to the fisheries business was estimated between 4 and 7 billion dollars (UNAVCO, 2010).

The 2011 Tohoku earthquake, often referred to as the Great East Japan Earthquake, had epicentre off the Pacific coast of Japan with a 9.0 magnitude, producing a tsunami 10 meters high along Japan's north-eastern coast (Gihooly, 2011). **Figure 2** shows the Tohoku earthquake observed tsunami heights. The wave caused widespread devastation, with an official count of 18.550 people confirmed to be killed/missing. The highest tsunami reached a total height of 40.5 metres. In addition the tsunami precipitated multiple hydrogen explosions and nuclear meltdown at the Fukushima I Nuclear Power Plant.

Clearly in such catastrophic events it is crucial to rapidly recognize the occurrence of tsunamis in order to promptly alert local communities to minimize human losses and to respond adequately providing the required humanitarian aid. However, a tsunami cannot be precisely predicted, even if the magnitude and location of an underwater earthquake is known, because it may depend on many different factors. However, there are some automated systems, which can provide warnings immediately after an earthquake in time to save lives. These systems rely on the assumption that a tsunami is defined as – and can be detected from – an unusually large change in the mean sea surface slope. One of the most successful systems uses bottom pressure sensors, attached to buoys, which constantly monitor the pressure of the overlying water column and provide the current tide level at the specific observing point. To recognize a tsunami the system needs to forecast the next occurring tide signal and to compare it with the observed tide signal in the same time interval. If the difference between the two signals overcomes a certain threshold, and if this event occurs several times consecutively, than it is likely that a tsunami is occurring in the observed zone and the tsunami warning is triggered, by initiating also the required response measures. Tidal analysis and forecasting are then crucial problems to face in order to provide with a high probability a correct tsunami warning.

Deep-sea measurements are perhaps the main means of detecting tsunamis generated either by earthquakes or by submarine landslides. They are collected at a standard sampling rate of 15 s by bottom pressure recorders (BPRs) located at water depths ranging from hundreds to some thousands of meters. It is at such water depths that BPRs can exclusively detect pressure fluctuations induced by propagating waves within the tsunami and tidal frequency band. Pressure fluctuations are, in fact, negligible at water depths greater than half a wavelength. This makes deep sea an ideal filter for waves characterized by lower periods (i.e. an ideal low-pass filter).

BPRs measure pressure fluctuations indirectly by evaluating the vibrational variations that these fluctuations cause in a quartz-bean piezometrically induced to oscillate in its lowest resonant flexural mode. Although the idea of correlating vibrational frequency with pressure-induced mechanical motion dates back to the end of the 1960s (Eble and Gonzlez, 1991), BPR technology has received a strong boost during the last 20 years. Such developments occurred under the Deep-ocean Assessment and Reporting of Tsunamis (DART) program (Milburn et al., 1996) run by the Pacific Marine Environmental Laboratory (PMEL) within the U.S. National Oceanic and Atmospheric Administration (NOAA).

Optimal use of BPR measurements depends both on instrument location and on the effectiveness of the detection algorithm. Whilst the instrument location should be determined by a tsunami hazard assessment (i.e. a previous knowledge both of probable tsunami sources and of places at risk along the coast), an effective forecasting algorithm should fast, capable to identify the waveform of a tsunami, and able to discriminate a tsunami from other sea-level oscillations that, falling within the tsunami and tidal frequency band, are 'disturbances' in the context of tsunami detection.

The main "disturbance" recorded by a BPR is caused by the superposition of actual sea-surface fluctuations (e.g. planetary waves, astronomical and meteorological tides or gravitational normal modes) and background sea noise. The close prediction of such a "disturbance" makes it possible to filter it out simply by subtracting the values observed from those predicted. The actual propagation of a tsunami can then be monitored by checking the amplitude of the filtered signal against a prescribed threshold. The amplitude of a perfectly filtered signal should, in fact, be equal to zero in the absence of a propagating tsunami.

Furthermore, the closer the prediction (i.e. the better the filtering performance), the lower the threshold to be prescribed, and the smaller the detectable tsunami. The core of the algorithm is therefore its method of using preceding actual observations to make (and update) predictions; in other words, its method of extrapolating new data from past observations.

The aim of this paper is to provide the communities of both researchers and practitioners with a broadly applicable, up to date coverage of tidal analysis and forecasting methodologies that have proven to be successful in a variety of circumstances, and that hold particular promise for success in the future. Classical and novel methods will be reviewed in a systematic and consistent way, outlining their main concepts and components, similarities and differences, advantages and disadvantages.

The present paper is structured as follows. The literature on tidal analysis and forecasting methods is presented in Section 2. Section 3 reports the details of conventional tidal forecasting methods that are based on harmonic analysis. These traditional harmonic methods rely on models based on the analysis of astronomical components and they can be inadequate when the contribution of non-astronomical components, such as the weather, is significant. The next sections contain other alternative approaches which have been developed in the literature and that are able to deal with these situations. In particular Section 4 describes standard high or band pass filtering techniques including, among the others, Kalman filters. However, the relatively deterministic character and large amplitude of tidal signals make special techniques, like wavelets transform analysis methods (Section 5) and artificial neural networks (Section 6), more effective. Conclusions and suggestions for further research are reported in Section 7.

# 2  Literature review

An early explanation of tides was given by Galileo Galilei in his 1616 Dialogue on the Tides. However, the resulting theory was incorrect; he attributed the tides to water sloshing due to the Earth's movement around the sun, hoping to provide mechanical proof of the Earth's movement. At the same time Johannes Kepler correctly suggested that the Moon caused the tides, based upon ancient observation and correlations, an explanation which was rejected by Galileo.

Isaac Newton was the first person to explain tides by the gravitational attraction of masses. His explanation of the tides (and many other phenomena) was published in the Principia (1687) and used his theory of universal gravitation to account for the tide-generating forces as due to the lunar and solar attractions. Newton and others before Pierre-Simon Laplace worked with an equilibrium theory, largely concerned with an approximation that describes the tides that would occur in a non-inertial ocean evenly covering the whole Earth (Lisitzin, 1974). The tide-generating force (or its corresponding potential) is still relevant to tidal theory, but as an intermediate quantity rather than as a final result; theory has to consider also the Earth's accumulated dynamic tidal response to the force, a response that is influenced by bathymetry, Earth's rotation, and other factors (Wahr, 1995).

In 1740, the Académie Royale des Sciences in Paris offered a prize for the best theoretical essay on tides. Daniel Bernoulli, Leonhard Euler, Colin Maclaurin and Antoine Cavalleri shared the prize (Cartwright, 1999). Maclaurin used Newton's theory to show that a smooth sphere covered by a sufficiently deep ocean under the tidal force of a single deforming body is a prolate spheroid (essentially a three dimensional oval) with major axis directed toward the deforming body. Maclaurin was the first to write about the Earth's rotational effects on motion. Euler realized that the tidal force's horizontal component (more than the vertical) drives the tide. In 1744 Jean le Rond d'Alembert studied tidal equations for the atmosphere which did not include rotation.

Pierre-Simon Laplace formulated a system of partial differential equations relating the ocean's horizontal flow to its surface height, the first major dynamic theory for water tides. The Laplace tidal equations are still in use today. William Thomson, 1st Baron Kelvin, rewrote Laplace's equations in terms of vorticity which allowed for solutions describing tidally driven coastally trapped waves, known as Kelvin waves (Cartwright, 1999).

Laplace's improvements in theory were substantial, but they still left prediction in an approximate state. This position changed in the 1860s when the local circumstances of tidal phenomena were more fully brought into account by William Thomson's application of Fourier analysis to the tidal motions. Thomson's work in this field was then further developed and extended by George Darwin: Darwin's work was based on the lunar theory current in his time. Darwin can be consider the father of classical tidal harmonic analysis, in which the tidal forcing is modeled as a set of spectral lines, i.e., the sum of a finite set of sinusoids at specific frequencies, referred to as constituents. His symbols for the tidal harmonic constituents are still used. He also first formulated a practical useful least-squares method for tidal harmonic analysis. Darwin's harmonic developments of the tide-generating forces were later brought up to date with modern developments by Arthur Thomas Doodson whose development of the tide generating potential (TGP) in harmonic form was carried out and published in (Doodson, 1921). He perfected Kelvin's theory by elaborating the formal treatment of the slowest astronomical periodicities. Inspired by the then-latest lunar theory of E W Brown, Doodson distinguished 388 different tidal frequencies and devised a practical system for specifying the different harmonic components of the tide-generating potential, the Doodson Numbers, a system still in use today to denote the various tidal species and constituents.

Since the mid-twentieth century further analysis has generated many more terms than Doodson's 388. About 62 constituents are of sufficient size to be considered for possible use in marine tide prediction, but sometimes many less even than that can predict tides with useful accuracy. The following are among the major tidal constituents contributing to the astronomical tide:
- M2 - Principal lunar semidiurnal constituent (speed: 28.984 degrees per mean solar hour);
- S2 – Principal solar semidiurnal constituent (speed: 30.000 degrees per mean solar hour);
- N2 - Larger Lunar elliptic semidiurnal constituent (speed: 28.440 degrees per mean solar hour);
- K1 - Luni-solar declinational diurnal constituent (speed: 15.041 degrees per mean solar hour);
- O1 - Lunar declinational diurnal constituent (speed: 13.943 degrees per mean solar hour);
- M4 - First overtide of M2 constituent (speed: 2 x M2 speed);
- M6 - Second overtide of M2 constituent (speed: 3 x M2 speed);
- S4 - First overtide of S2 constituent (speed: 2 x S2 speed);
- MS4 - A compound tide of M2 and S2 (speed: M2 + S2 speed).

Munk and Cartwright (1966) reformulated the tidal problem in terms of admittances to account for the presence of a continuous spectrum of background noise (the so-called response method). This work, however, resulted in little change in practical tidal analysis procedures. Aside from the technical refinements introduced by Doodson (1921), tidal harmonic analysis as it is used in practice has remained nearly static in the next years. The improvements that have occurred since 1921 have been related to removal of the effects of minor constituents that cannot be determined from a one-year record, inference from one station to another, more precise specification of astronomical inputs, treatment of unevenly spaced data, treatment of vector data, and development of numerically efficient software (e.g. (Godin, 1972) and (Foreman, 1977)). The calculations of tide predictions using the harmonic constituents are laborious, and until about the 1960s they were carried out by using a mechanical tide-predicting machine, a special-purpose form of analog computer now superseded by digital electronic computers.

In recent years several systems for automatic, real-time tsunami detection in sea-level measurements have been developed by the research community in order to promptly alert the costal population whenever such a disastrous event is occurring. Direct detection in these systems, also referred to as Tsunami Early Warning Systems (TEWS), is essential in order to confirm the actual generation and propagation of a tsunami, i.e. to upgrade or cancel the rapid initial warning usually given on the sole basis of seismic data. Whatever the sensor type and location, the optimal use of sea-level measurements depends on the effectiveness of the detection algorithm implemented in the software of the sensor. The requisites for an effective TEWS algorithm have been identified as the following (Beltrami, 2008):
    1. an ability to discriminate a tsunami from other sea-level oscillations that, falling within the frequency band detected by the sensor, are "disturbances" in the context of tsunami detection;
    2. an ability to identify the waveform of a tsunami and to characterize it in terms of both amplitude and period;
    3. a fast computational time.

As far as requisite (1) is concerned, it should be noted that, although the presence of several frequencies in a sensor recorded signal guarantees the instrument's poli-functionality, only filtering out non tsunami waves makes it possible to monitor the actual propagation of a tsunami by checking either the amplitude or the slope of the filtered signal against a prescribed threshold (Mofjeld, 1997; Beltrami, 2008). Therefore some form of low-pass digital filters has to be performed in a tsunami detection algorithm in order to filter out the "disturbance" recorded by the sensors.

McGehee and McKinney (1995) and (Leva, 2004) developed two different real-time tsunami detection algorithms that are used by the Italian Civil Protection within the Stromboli's TEWS. Both these algorithms are based on time-domain point-to-point Moving-Average (MA) filters, also known as running-mean filters. A further algorithm (Shimizu et al., 2006), that relies on a Finite Impulse Response (FIR) time domain filter, has been more recently proposed by the Port and Airport Research Institute of Japan (PARI) for the automatic, real-time detection of a tsunami in the sea-level measurements collected by the Dopplertyped Wave Directional Meter (DWDM) and the GPS buoys of the Japanese Nationwide Ocean Wave information network for Ports and HArbourS (NOWPHAS). Furthermore, Beltrami and Di Risio (2011) developed a new tsunami detection algorithm mainly based on an Infinite Impulse Response (IIR) time domain filter which is able to meet thoroughly the requisites set out above (see also (Bressan and Tinti, 2012)).

As for filtering methods, a wide application also of Kalman filters (Kalman, 1960) to tidal analysis and forecasting is found in the literature. The first application of the Kalman filter in the context of numerical weather prediction was first presented by Ghil et al. (1981). Miller and Cane (1989) then applied the Kalman filter for a simple model of sea level anomalies in the tropical Pacific, using tide gauge data from six selected island stations to update the model. Their Kalman filter required detailed statistical assumptions about the errors in the model and the data. The error model in their filter was a simple covariance function with parameters fit from the observed differences between the tide gauge data and the model output. The calibrated error model was used in the Kalman filter to generate monthly sea level height anomaly maps. The filtered maps exhibited fine structure that is absent from the unfiltered model output, even in regions removed from the data insertion points. Given that only six data points were embedded in the data assimilation, the results obtained by the Kalman filter of Miller and Cane (1989) were quite encouraging, and they were further refined by the Kalman filtering method for the tropical Pacific in (Chan et al., 1996) and for El Niño-Southern Oscillation in (Chen and Cane, 2008).

Later Yen et al. (1996) proposed a short-term tide level prediction method based on Kalman filtering using just few days of tide measurements of the Kaohsiung Harbor in Taiwan. In general accurate predictions of tide levels could not be obtained by conventional methods without a long length (about one month or more) of tide measurements. In the method proposed by the authors, a harmonic tide level model is used to predict tide levels. Then the parameters of the tide level model (i.e. the amplitudes of the harmonic components) are estimated by Kalman filtering technique using just a few-day tide record with the assumption of known angular frequencies. Their predicted tide levels were in good agreement with the observed real data.

Successively other tide analysis and forecasting methods based on Kalman filtering were largely proposed in the literature. The most relevant perhaps is the contribution by Choi et al. (2000), in which sea-level predictions from short observation record, based on harmonic tidal models with their parameters estimated by the Kalman filter algorithm, were tested for Macau. Afterwards

Sørensen et al. (2006) presented a parameter sensitivity study of three well known general Kalman filter approaches (the ensemble Kalman filter, the reduced rank square root Kalman filter, and the steady Kalman filter) to use for the assimilation of water levels in a three dimensional hydrodynamic modelling system. Their sensitivity study further demonstrates the effectiveness of Kalman filtering in dealing with this kind of systems.

As it has been discussed, numerous models for tidal analysis and forecasting have been carried out in the literature. The accuracy of harmonic models depends entirely on accurate observed data over a long-term tidal record, which is used to determine the parameters of the tidal constituents. This is the major shortcoming of the harmonic models. The Kalman filtering method has been applied to determine harmonic parameters with a limited amount of tidal measured data (Altaf et al., 2013). However, the model is only applicable for short-term prediction, rather than long-term prediction (e.g. Yen et al. (1996)). In addition the use of harmonic analysis may be inadequate when tidal interactions with varying topography can produce large internal waves and bores whose characteristics are highly sensitive to ambient stratification. In such cases other techniques such as wavelet analysis have been proposed in the literature (Flinchem and Jay, 2000; Sun et al., 2006; Zhong and Oyadiji, 2013).

When data for the observed periods are lost or incomplete, existing analytical methods like harmonic analysis and Kalman filtering are not effective in supplementing the lost data. Therefore, in such cases it is important to find an accurate tidal level prediction technique. For this reason, recently, artificial neural networks (ANNs) have been used in the literature as an alternative tidal level forecasting approach. Based on limited field data, the neural network method can predict hourly, daily, weekly or monthly tidal level more accurately than, for example, harmonic analysis methods. Vaziri (1997) compared the ability of an artificial neural network with multiplicative autoregressive integrated moving average modelling. Deo and Chaudhari (1998) used the neural network model to produce the tidal curves of a subordinate station based on the data of a standard or reference tidal station.

Neural networks were first applied to tide forecasting by using the field data of both diurnal and semidiurnal tides by Tsai and Lee (1999). However, their model was only applicable for the prediction of diurnal and semi-diurnal tides. In fact, mixed tides are more likely to occur in the field than diurnal and semi-diurnal tides (Lee et al., 1998). Lee et al. (2002) and Lee (2004) applied a neural network to predict different types of tides and found that the technique can be effective. However, their methods depend on harmonic parameters and cannot predict non-astronomical tidal level. Huang et al. (2003) developed a regional neural network water level model based on long-term tidal data from a NOAA station. Even though the non-astronomical components effect was significant, the model's prediction accuracy was good. Cox et al. (2002) and Lee (2006) addressed the prediction of tidal level including non-astronomical components. Lee and Jeng (2002) developed an advanced ANN model for tide forecasting using a short-term tidal record and considering all diurnal, semi-diurnal and mixed tides in their model. The data from three harbours in Taiwan were used as case studies, and the effects of the neural network structure, including training techniques and learning algorithms, were discussed in detail.

Beltrami (2008; 2011) implemented algorithms based on the use of artificial neural networks in the software of bottom pressure recorders (BPRs) for the automatic, real-time detection of a tsunami within recorded signals. He compared his algorithms for tsunami detection to the one developed under the Deep-ocean Assessment and Reporting of Tsunamis (DART) program run by Mofjeld (1997) for the U.S. National Oceanic and Atmospheric Administration (NOAA). His results showed a consistent improvement in detection performance by using the ANN approach. Other recent works on the subject include (Liang et al., 2008; Deo, 2010; Wang et al., 2012).

# 3 Harmonic analysis methods

Classical harmonic analysis methods model the tidal forcing as a set of spectral lines, representing the sum of a finite set of sinusoids at specific frequencies. These frequencies are obtained as combinations of sums and differences of integer multiples of six fundamental frequencies, called Doodson Numbers (Doodson, 1921), which arise from planetary motions (Godin, 1972). These fundamental parameters represent effects of rotation of the earth, the orbit of the moon around the earth and the earth around the sun, and periodicities in the location of lunar perigee, lunar orbital tilt, and the location of perihelion (Pawlowicz, 2002).

In Astronomy an *equilibrium response* is the steady state response of the climate system (or a climate model) to an imposed radiative forcing. The phase and amplitude of an equilibrium response can be predicted by the values observed if the response of the earth was fast enough that the surface deformation was effectively in equilibrium with the forcing at all times (McGuffie and Henderson-Sellers, 2005). Although oceans are not in equilibrium with tidal forcing, in general tidal amplitudes are small compared with the total ocean depth yielding to nearly linear dynamics. This implies that the forced equilibrium response contains only those frequencies present in the tidal forcing (McGuffie and Henderson-Sellers, 2005).

To determine the relative phase and amplitude of each frequency in the tidal response, harmonic analysis methods make use of least-squares fit. The phase and amplitude of each frequency represent a compression of the data in the complete tidal time series. This data can be compared with similar data at other locations to understand the characteristics of tidal dynamics, or can be used to synthesize time series of tidal effects at other times for predictive purposes.

The rest of this section is structured as follows. In Section 3.1, the form of the equilibrium potential is described, whereas Section 3.2 includes the mathematical basis of the harmonic technique to estimate phase and amplitude of tidal frequencies. Finally Section 3.3 reports the main drawbacks of classical harmonic analysis methods.

## 3.1 Tidal potential

The effect of gravitational force vectors from the sun and moon, **F**, can be written as the gradient of a scalar potential *V*:

$$\boldsymbol{F} = -\nabla V. \qquad (I)$$

The magnitude of the potential at the earth's surface at any time is a function of lunar time $\tau$ (defined to begin at ''lunar midnight'') and other time-dependent astronomical variables (Doodson, 1921; Foreman and Henry, 1989):
- *s*, which is the mean longitude of moon;
- *h*, which is the mean longitude of sun;
- *p*, which is the longitude of perigee;
- *N'*, which is the negative of the longitude of the ascending node;
- *P'*, which is the longitude of perihelion.

These terms, all expressed in units of cycles, are easily obtained by Seidelmann's formulas (Seidelmann, 1992), and they are combined with the Doodson numbers (*i', j', k', l', m', n'*) for a particular constituent frequency, σ, into the astronomical argument $V_\sigma$ as follows:

$$V_\sigma = i' \cdot \tau + j' \cdot s + k' \cdot h + l' \cdot p + m' \cdot N' + n' \cdot p'. \qquad (II)$$

From the astronomical argument $V_\sigma$ the constituent frequency σ is given as (Song et al., 2011):

$$\sigma = 2\pi \frac{dV_\sigma}{dt}. \qquad (III)$$

Frequencies with common *i', j',* and *k'* terms form a *subgroup*, while frequencies with a common *i'* number are referred to as a common *species*. For example the species with *i' = 0* is called *slow species*; with *i' = 1* is called *diurnal species*; *i' = 2* is called *semidiurnal species*.

Considering the astronomical argument $V_\sigma$, the tidal potential *V* can be expressed as (Godin, 1972):

$$V = \sum_{i'=1}^{3} G_{i'}(\theta) \cdot \sum_{j',k',l',m',n'} A'_{i',j',k',l',m',n'} \cdot \cos(2\pi V_\sigma) + G'_{i'}(\theta) \;\cdot$$

$$\cdot \sum_{j',k',l',m',n'} B'_{i',j',k',l',m',n'} \cdot \sin(2\pi V_\sigma) \qquad (IV)$$

where:

- $G_{i'}(\theta)$ and $G'_{i'}(\theta)$ are geodetic functions which depend on species *i'*, the latitude, *θ*, and also on some constants such as the radius of the earth and the masses and separations of the earth, moon, and sun;

- $A'_{i',j',k',l',m',n'}$ and $B'_{i',j',k',l',m',n'}$ are tabulated constant values and, for a given Doodson number set either one among the two values is nonzero, but not both.

The equilibrium amplitude for a particular constituent, generated for a particular latitude *θ*, is defined as (Godin, 1972; Song et al., 2011):

$$g^{-1} \cdot G_{i'}(\theta) \cdot A'_{i',j',k',l',m',n'} \qquad (V)$$

or either as:

$$g^{-1} \cdot G'_{i'}(\theta) \cdot B'_{i',j',k',l',m',n'} \qquad (VI)$$

where *g* is the gravitational acceleration.

## 3.2 Tidal response estimates

Consider a time series of observations of either real or complex numbers $y(t)$, with $t = t_1, t_2, ..., t_M$, arranged as a vector regularly spaced with respect to the time at an interval $\Delta t$ (note that missing observations may be marked by using a ''missing data'' symbol like NaN, the IEEE arithmetic representation for Not-a-Number).
The tidal response is modeled as (Pawlowicz, 2002; Foreman and Henry, 1989):

$$x(t) = b_0 + b_1 t + \sum_{k=1}^{N} a_k e^{i\sigma_k t} + a_{-k} e^{-i\sigma_k t}, \qquad (VII)$$

where N constituents (each with a unique Doodson number set) are used. Each constituent has a frequency $\sigma_k$ which is known from the development of the potential (see Equation (III) and Equation (IV)), and a complex amplitude $a_k$ which is unknown, even if $y(t)$ is a real time series then $a_k$ and $a_{-k}$ are complex conjugates. The first two terms $b_0$ and $b_1 t$ are optional and are used to balance a possible offset of the time series.
The traditional approach to represent the tidal response uses real sinusoids (Godin, 1972; Foreman, 1977):

$$x(t) = b_0 + b_1 t + \sum_{k=1}^{N} A_k \cos \sigma_k t + B_k \sin \sigma_k t, \qquad (VIII)$$

and is related to Equation (VII) by $A_k = a_k + a_{-k}$ and $B_k = i(a_k + a_{-k})$. This real representation of the tidal response as sum of sinusoids is more convenient for linear error analysis purpose.

The *N* constituents to consider in the series are selected from a set of astronomical and shallow-water constituents by an automated selection algorithm (Foreman, 1977; Foreman et al., 2009). This works by preselecting all astronomical and a subset of the most important shallow-water constituents (usually around 24), and then by ranking them in order of predefined importance based on equilibrium amplitudes. Those constituents whose frequencies are less than a Rayleigh resolution frequency limit $\alpha(N\Delta t)$, with default $\alpha = 1$, are then discarded but, if required, additional shallow-water constituents may be added to the set. If the relative phase/amplitude of two constituents that are otherwise unresolvable is known from other sources, then an inference procedure can be carried out (Pawlowicz, 2002; Nidzieko, 2010). Alternatively, constituent lists can be explicitly specified.

At this point to determine the relative phase and amplitude of each frequency in the tidal response *x(t)*, a least-squares fit is used. Phase and amplitude of each constituent are then the coefficients which minimize the following function (Foreman, 1977; Foreman et al., 2009):

$$\min \sum_{m=1}^{M} |x(t_m) - y(t_m)|, \qquad (IX)$$

which can be also expressed in a more formal and compact way as (Nidzieko, 2010):

$$\min \|TC - Y\|, \qquad (IX)$$

where $Y = [y(t_1), y(t_2), \ldots, y(t_M)]'$ is the array of the observed tide signal, $C = [b_0, b_1, a_1, a_{-1}, a_2, a_{-2}, \ldots, a_N, a_{-N}]'$ is the array of the unknown coefficients to determine by the fit, and $T$ is an [M x 2N+2] matrix of linear and sinusoidal basis functions evaluated at observation times and derived by Equations (VII)-(VIII).

The result of the least-squares fit produces a set of complex pairs $\{a_k, a_{-k}\}$ for each constituent $k$, which are also possible to correct for a larger precision, if latitude is specified, by a so-called *nodal (or satellite) modulation* (Pawlowicz, 2002). Consider a main peak of index $k$ with satellites with indices $kj$. The effect of the different satellites will be to slowly modulate the phase/amplitude of the main peak over various periods. The results of the fitted least-squares fit can be corrected then by modifying the real response of the main constituent $a_k$. In particular the presence of the satellites affects the amplitude of $a_k$ by a factor $f_k$ referred to as *nodal correction amplitude*, and the phase of $a_k$ by an angle $u_k$ referred to as *nodal correction phase*, as specified in the following equation (Pawlowicz, 2002; Nidzieko, 2010):

$$\hat{a}_k e^{i\sigma_k t} = f_k a_k e^{i\sigma_k t + iu_k} = a_k e^{i\sigma_k t} + \sum_j a_{kj} e^{i\sigma_{kj} t}, \qquad (X)$$

where $\hat{a}_k$ is the corrected response coefficient for the main peak constituent $k$. Simplifying by the common term $a_k e^{i\sigma_k t}$ the expression becomes:

$$f_k e^{iu_k} = 1 + \sum_j \frac{a_{kj}}{a_k} e^{i(\sigma_{kj} - \sigma_k)t} \approx 1 + \sum_j \frac{a_{kj}}{a_k}, \qquad (XI)$$

as long as $(\sigma_{kj} - \sigma_k)t$ remains a small number (experimentally by $N\delta t < 8$ years).

In general the true phases and amplitudes of the satellites are not known. However, since their frequencies are very similar to that of the main peak $k$ it is standard to assume that the ratio of true amplitudes is the same as the ratio of amplitudes in the equilibrium response, and the difference in true phases will be equal to the difference in equilibrium phases (Foreman et al., 2009). The nodal corrections are thus computed from the equilibrium response (Equation (IV)).

Once the coefficients $\{a_k, a_{-k}\}$ for each constituent $k$ have been evaluated they can be expressed in a standard way by describing an ellipse as follows (Pawlowicz, 2002; Foreman et al., 2009):

$$\begin{cases} L_k = |a_k| + |a_{-k}|, \\ l_k = |a_k| - |a_{-k}|, \\ \theta_k = \frac{ang(a_k) + ang(a_{-k})}{2} mod(180), \\ g_k = v_k - ang(a_k) + \theta_k. \end{cases} \qquad (XII)$$

where $L_k$ and $l_k$ represent the length of the semi-major and semi-minor axis of the ellipse, $\theta_k$ the inclination of the northern semi-major axis counter-clockwise from due east, and $g_k$ the phase of the constituent response, also referred to as *Greenwich phase*, i.e. the phase referenced to the phase of the equilibrium response at 0° longitude (the Greenwich meridian). This can be interpreted as reporting the phase of the response at the time when the equilibrium forcing is at its largest positive value at 0° longitude. In this way it is simplest to find the fitted phase at the central time of the record ($t = 0$); the equilibrium phase $v_k$ is then just $V_a$ for the given constituent computed at the Julian date corresponding to this central time, with possible adjustments of ¼, ½, or ¾ cycle depending on whether A or B is non-zero, and their signs (Pawlowicz, 2002). If $l_k > 0$ (or $< 0$) then the ellipse is traced in a counter-clockwise (or clockwise) direction. In the case of scalar time series, the parameter $L_k$ is the amplitude of the ellipse, and $l_k = \theta_k = 0$ (i.e. the ellipse degenerates to a line along the positive axis). The representation least-squares fit resulting parameters as features of an ellipse may be helpful for further analysis and tide prediction (Nidzieko, 2010; Matte et al., 2013; Moftakhari et al. 2013).

## 3.3 Main drawbacks of harmonic analysis

The following list enumerates the main drawbacks of classical harmonic analysis. For further discussions and more comprehensive descriptions on the limitations of harmonic analysis techniques, the reader is referred to (Godin, 1991; Foreman and Henry, 1989; Pawlowicz, 2002):

1. The first concern is with respect to the length of the available tidal record. In harmonic analysis the modulation of perihelion is disregarded, as it is nearly constant over historical time. This implies that to obtain exactly all of the listed frequencies by harmonic analysis, a time record of around 18 years should be available (Foreman et al., 1995). However in practice record lengths are often 1 year or shorter and, in these cases, the tidal signals are sinusoids whose phase and amplitude change slowly with time, and that can be considered effectively constant for these short record lengths (Foreman and Henry, 1989). This simplification comes from the assumption that the phase/amplitudes of response sinusoids with similar frequencies (i.e., those whose first three Doodson numbers are identical) are in the same proportion as those of the equilibrium response under the reasonable premise that the ocean response should be similar at similar frequencies. In such a cluster, large equilibrium peaks are surrounded by small subsidiary peaks in frequency-space which provide the nodal (or satellite) modulations to the main peak, as reported in Section 3.1.

2. Another problem arises when dealing with shorter record lengths. The frequency resolution further degrades until even dissimilar constituents are unresolvable. In this case the best solution to find the absolute phase/amplitude of dissimilar constituents consists of applying *inference*. However in order to apply this technique it is required to know the relative differences in phase/amplitude between the two unresolved constituents from other nearby data. If this is not possible to apply inference then the only way is to either discard the smaller constituents and fit only to the largest in a given frequency interval, or to use the equilibrium response to establish the desired differences (Pugh, 1987; Nidzieko, 2010; Matte et al., 2013).

3. A further drawback of classical harmonic analysis is that the degree to which a given constituent represents true tidal energy as opposed to the energy of a broad-band non-tidal process is not determined; the resulting least-squares fit in harmonic analysis is likely to methods in the produced time series of both tidal and non-tidal signals. But harmonic analysis needs to perceive the two different signal sources to have a better estimate of the tidal behaviour, and to allow quantitatively the comparison of the two different analyses. However in harmonic analysis an immediate method to evaluate deterministically whether the resulting phase/amplitude of a given sinusoid is meaningful or not does not exist. To address this issue, Munk and Cartwright (1966) proposed the *response method* as an alternative harmonic approach, although it has not found widespread use in the literature. Instead it has been preferred to include in harmonic methods some kind of confidence intervals to differentiate between true tidal frequencies and broad-spectrum variability. There are two steps to producing confidence intervals. First, it is necessary to estimate the characteristics of non-tidal or residual noise affecting the parameters $a_k$ (or $A_k$, $B_k$). Second, these estimates are converted into confidence intervals for the standard parameters through a nonlinear mapping. For more details the reader is referred to (Pawlowicz, 2002; Foreman et al., 2009). Recently, Leffler and Jay (2009) presented a robust fitting to extend the ordinary least squares calculation of harmonic analysis in order to be more resistant to broad-spectrum noise. Since the variance of the amplitude and phase is calculated from the power spectrum of the residual, their method filters broad-spectrum noise and reduces the residual variance, increasing the confidence of the computed parameters and allowing more low-amplitude constituents to be resolved from the background noise (Leffler and Jay, 2009).

4. Further problems with classical harmonic analysis arise in coastal regions where the tidal response is in the form of a wave propagating onshore (Godin, 1991; Moftakhari et al. 2013). In large estuaries, the seasonal change in salinity and flow may change the dynamic response but as these changes can vary from year to year the tidal process is not really stationary. Instead spectral peaks are broadened so that they are no longer pure lines but, depending on the situation, such variations may be treated as lines in the analysis. Within smaller estuaries, tidal height variations may be significant compared to water column depth and a variety of non-linear effects can occur. For example, flood periods shorten and intensify and ebbs lengthen. As long as these effects are reasonably deterministic they may be handled by adding extra "shallow water"' constituents which occur at sum/difference frequencies of the major constituents (Godin, 1991; Moftakhari et al. 2013). More problematic in these regions are the effects of internal variability. Tidal interactions with varying topography can produce large internal waves and bores whose characteristics are highly sensitive to ambient stratification (Matte et al., 2013). In such cases the use of harmonic analysis becomes inadequate and, as it will be shown in the next sections, other techniques such as wavelet analysis or neural networks have been proposed in the literature.

# 4   Sea-level filters

*Mean sea level* (MSL) is a measure of the average height of the ocean's surface (such as the halfway point between the mean high tide and the mean low tide), used as a standard in reckoning land elevation. The concept of a "mean sea level" is in itself rather artificial, because it is not possible to determine a figure for mean sea level for the entire planet, and it varies quite a lot even on a much smaller scale. This is because the sea is in constant motion, affected by the high and low-pressure zones above it, the tides, local gravitational differences, and so forth. The best one can do is to pick a spot and calculate the mean sea level at that point and use it as a datum.

It is now normal practice for national authorities to calculate monthly and annual mean sea levels from periodic (usually between 1 minute and 1 hour) values of observed sea level. The basis for any scientific analysis of sea level relies then on the analysis of long time series of careful measurements. Any instantaneous measurement of sea level in a series may be considered the sum of three component parts:

$$\text{observed level} = \text{mean sea level} + \text{tide} + \text{meteorological residuals}$$

Each of these component parts is controlled by separate physical processes and the variations of each part are essentially independent of the variations in the other parts. An acceptable set of definitions is (UNESCO, 1985):

- Tides are the periodic movements of the seas, which have coherent amplitude and phase relationship to some periodic geophysical force. The dominant forcing is the variation in the gravitational field on the surface of the earth due to the regular movements of the earth-moon and earth-sun systems. These cause gravitational tides. There are also weak tides generated by periodic variations of atmospheric pressure and on-shore offshore winds that are called meteorological tides.

- Meteorological residuals are the non-tidal components, which remain after removing the tides by analysis. They are irregular, as are the variations in the weather. Sometimes the term surge residual is used but more commonly surge is used to describe a particular event during which a very large non-tidal component is generated.

- Mean sea level is the average level of the sea, usually based on hourly values taken over a period of at least a year. For geodetic purposes the mean level may be taken over several years. The frequency with which different observed hourly levels occur over a long period of observation has a definite pattern. More elaborate techniques of analysis allow the energy in the sea level variations to be split into a series of frequency or spectral components. The main concentration of energy is in the semidiurnal and diurnal tidal bands, but there is a continual background of meteorological energy, which becomes more important for longer periods or lower frequencies.

When a Tsunami occurs in a coastal region, mean sea level changes occur by a vertical crustal movement, which displaces the seabed. The tsunami wave characteristics will depend on the amplitude of the displacement and the dimensions of the seabed involved. As the wave approaches shallow coastal waters its amplitude increases and there are multiple reflections and refractions, which combine to give very large local amplitudes. Mean sea level monitoring is in place in exposed coastal regions for the operational use of sea level data in particular for tsunami warning (or in general other natural phenomena like storm surges) by means of some automatic procedures (Holgate et al., 2013). In these circumstances raw data are normally registered at time intervals between 1 minute and 1 hour (the most common being 5, 6 and 10 minutes), and rapid warning transmissions are in place in case of an occurring event for immediate alerting of human population.

Apart from keeping higher frequency signals for other purposes, it is always necessary to obtain filtered hourly values for further sea level data processing. The filtering process will eliminate higher frequencies dependent on the frequency cut-off. The simplest way is to take the arithmetic average, but more elaborate methods include the application of low-pass numerical filters to eliminate tides and surges before taking the average (Pugh, 1987; Holgate et al., 2013). It is well known that an effective low-pass digital filter should possess five, often mutually exclusive, essential qualities (Beltrami and Di Risio, 2011):
1. a sharp cut-off, so that unwanted high-frequency components are effectively removed;
2. a comparatively flat pass-band that leaves the low frequencies unchanged;
3. a clean transient response so that rapid changes in the signal do not result in spurious oscillation or 'ringing' within the filtered record;
4. a zero phase shift;
5. an acceptable computational time.

In the context of automatic, real-time tsunami detection, a further requisite should also be met (Bressan and Tinti, 2012). The filter to be used should belong to the class of *causal or physically realizable filters*, i.e. to the class of filters that use as input only actually available or 'past' signal samples. It is to be noticed that the use of only 'past' samples makes a zero-phase response not possible for causal filters. A way to get round this problem is to design a filter that has, at least, a linear phase response, i.e. a symmetrical impulse response with location of symmetry shifted from the sample to be filtered. Indeed, since this shift does nothing but produce an identical shift in the output signal, a linear phase filter is equivalent to a zero phase one for most purposes (Shenoi, 2006). It is finally to be stressed that, in the context of real-time tsunami detection, the time-domain approach to filter design undoubtedly possess the appealing pro of making the filtered signal immediately available.

The rest of this section describes the main filtering methods that have been applied to the sea level data at short time intervals to obtain the main sea level signal. For more detailed descriptions of the filtering methods in the following sections the reader is referred to (Buades et al., 2005; Mitra and Kaiser, 1993; Shenoi, 2006; Harvey, 1989).

## 4.1 Arithmetic mean values

The most direct way of calculating monthly mean levels is to add together all the hourly values observed in the month, and then divide the total by the number of hours in the month. The annual mean level can be calculated from the sum of the monthly mean levels, weighted for the number of days in each month. Any days for which hourly values are missing should be excluded, but only these days are lost in the analysis. Small errors are introduced by the incomplete tidal cycle included at the end of the month. This method is used by many authorities because it requires little mathematical insight, yet produces values close to those obtained by more elaborate tide-eliminating techniques (Rossiter, 1958). If the averaging is done manually, the hourly levels can be tabulated in rows for hours and in columns for days; the monthly total may then be checked by summing each row and each column, and comparing the total for the rows with the total for the columns, although such laborious manual procedures are now seldom necessary (Pugh, 1987; Buades et al., 2005).

## 4.2 Point-to-point Moving-Average filtering methods

The point-to-point Moving-Average (MA) filter is a simple *Low Pass FIR* (*Finite-duration Impulse Response*) filter, commonly used for smoothening an array of sampled data (Shenoi, 2006). At time step $t_i$ this kind of filters expresses the output sample y($t_i$) as a weighted sum of the last *q* inputs:

$$y(t_i) = \sum_{k=0}^{q} a_k x(t_{i-k}). \qquad (XIII)$$

where *q* is the order of the filter, $x(t)$ and $y(t)$ respectively the original (i.e. to be filtered) and the filtered signal, and $a_k$ are the equation coefficients (which may vary for a FIR filter to another). Since Moving-Average filters do not use feedback, they are inherently stable. If the coefficients $a_k$ are symmetrical (in the usual case), then such filters are linear phases, so they delay signals of all frequencies equally (this is important in many applications; it is also straightforward to avoid overflow in an FIR filter (Mitra and Kaiser, 1993)).

The MA filter can be imagined as a window of a certain size (defined by the user) moving along the array, one element at a time. The middle element of the window is replaced with the average of all elements in the window. However, it is important to remember the value of new elements and not make the replacement until the window has passed. This must be done since all averages shall be based on the original data in the array. The main FIR methods applied for tide analysis and forecasting are described in the rest of this section.

### 4.2.1 Doodson $X_0$ filter

Belonging to the family of FIR filters is the *Doodson $X_0$ filter*, firstly proposed by Doodson (1921) and extended by Pugh (1987) along with other slightly different variations. The Doodson $X_0$ filter is a simple filter designed to damp out the main tidal frequencies. It is a symmetrical filter so no phase shifts are introduced in its mathematical model. Given the time-variant signal $X(t)$, the filtered value $X_F(t)$ at time *t* is computed (Pugh, 1987; Hasler and Hasler, 2007):

$$X_F(t) = F_0 \cdot X(t) + \sum_{m=1}^{M} F_m[X(t+m) + X(t-m)]. \qquad (XIV)$$

Then the computed $X_F(t)$ values must be normalized by selecting $M = 19$ sampled values both side of the central referring value, and then by performing a weighted average with the following weights (UNESCO, 1985):

$$\frac{(1010010110201102112\ 0\ 2112011020110100101)}{30}. \quad (XV)$$

If the initial values are at a higher frequency, they need to be first averaged to give at least hourly values. In some cases a slightly less accurate value of the monthly mean may be obtained by making a point-to-point average, especially when the labour of producing periodic levels is considered too great (Rossiter, 1961). This method of averaging, also called the $Z_0$ filter, is only appropriate when more accurate techniques cannot be used. However, the mean sea-level values obtained are usually reasonably close to the more accurate values (Pugh, 1987).

Other variants of the Doodson $X_0$ filter are discussed in (Godin, 1972), as well as in (Pugh, 1987) and (Hasler and Hasler, 2007). However their main mechanisms differ very slightly from the main model discussed in this paragraph, and they are very similar to each other.

### 4.2.2  FIR filter by McGehee and McKinney (1995)

The filtering algorithm by McGehee and McKinney (1995) used for tsunami detection works as follows. The filtered sea-level $\hat{\zeta}(t_i)$ is obtained by taking the arithmetic mean of the vector level $\{\zeta\} = \{\zeta(t_i), \ldots \zeta(t_{i-n+1})\}$ consisting of $n$ samples collected in the time interval defined by the duration $\Delta t$ of the averaging window. It is clear that $n$ is equal to $\Delta t \cdot f_s$, where $f_s$ is the sampling frequency. The averaging window is slid forward by an increment $s\Delta t_s = s/f_s$ (with $s$ an integer value), and the slope of the filtered sea-level between successive intervals is calculated. A tsunami will be triggered if the absolute value of the calculated slope exceeds a preselected threshold $TS_{slp}$ for some $m$ number of successive intervals $s\Delta t_s$. The filter on which the algorithm is based therefore belongs to the class of causal moving-average digital filters, which produce a linear phase alteration (McGehee and McKinney, 1995).

The choice of the values of the algorithm parameters is clearly site dependent, since it depends on both the climate disturbance wave and the tidal range at the sensor's location. The issue is clearly a trade-off between opposite demands. As pointed out by McGehee and McKinney (1995), the duration $\Delta t$ of the averaging window should be long enough to filter out the effects of meteorological disturbance on the tide mean. The longer this window duration is (and therefore the better the filtering performance), the higher the period of the shortest detectable tsunami. The lower bound of the band of the detectable-tsunami periods is, in-fact, determined by the value chosen for $\Delta t$. The upper bound depends on the amplitude of the minimum tsunami to be detected. Given this amplitude, the upper bound period should make the slope of the tsunami steeper than that of the steepest tidal wave measured at the location of interest. As far as the increment $s\Delta t_s$ is concerned, too short a value may lead to an immeasurable change in the slope between successive windows for an event of interest, while too long a value may lead to a warning that is not usable (i.e. a late warning). Finally, the threshold $TS_{slp}$ should be small enough to detect the minimum tsunami considered a threat, and large enough to reject usual changes in slope, as well as unusual but not-threatening ones.

The algorithm by McGehee and McKinney (1995) relies on the assumptions that: (i) the range of the maximum peak period of the tidal waves observed at the generic location of interest is equal to 18–60 s (the duration $\Delta t$ is therefore on the order of 10 times this period); (ii) the range of the semi-diurnal tide at the same location is on the order of 10 $m$; and (iii) the amplitude of the minimum-tsunami to be detected is equal to 0.5 *m*. These assumptions constrain the period of the tsunami to be detected to the band 10–40 *min*, i.e. to the band typical for tsunamis measured by a near-shore gage in the mid Pacific Ocean. It is to be noticed that nothing inherent in the algorithm prevents selecting different values of the algorithm's parameters. Therefore, in principle, these values can be selected in order to detect tsunamis characterized by a different period band.

Finally, the algorithm does not guarantee the characterization of the detected tsunami, particularly in terms of amplitude. Indeed, as already stated, the algorithm by McGehee and McKinney (1995) checks against a prescribed threshold the slope of the filtered signal and not its amplitude. In this regard, the algorithm belongs to the class of the *slope-discriminating* ones.

### 4.2.3   Italian Civil Protection filtering by Leva (2004)

The algorithm used by the Italian Civil Protection within the Stromboli's TEWS is characterized by a cascade of filters belonging to the class of causal moving-average digital filters and producing a linear phase alteration (Leva, 2004). In particular, given the sensors sampling interval (0.5 s), it works as follows (see
Figure **3**).

A vector $\{\zeta\}$ consisting of the 120 samples collected during the preceding 60 *s* (i.e. $\{\zeta(t_{i-1}), \ldots \zeta(t_{i-120})\}$) is updated at each time step $t_i$. The arithmetic mean $\bar{\zeta}$ of this vector is computed and updated every 60 *s*. This arithmetic mean is subtracted from each new sample $\zeta(t_i)$ in order to filter out tidal oscillations from the signal. The result $\zeta'(t_i) = \zeta(t_i) - \bar{\zeta}$ of this subtraction is stored in the vector $\{\zeta'\} = \{\zeta'(t_i), \ldots \zeta'(t_{i-29})\}$, consisting of the 30 values stored during the last 15 *s*. The filtered sea-level $\hat{\zeta}(t_i)$ at each new time step $t_i$ is then obtained by taking the mean $\bar{\zeta}_i$ of $\{\zeta'\}$. The duration of the tidal wave filtering window is therefore equal to 15 *s*. A tsunami will be triggered if $|\hat{\zeta}(t_i)|$ exceeds a preselected threshold $TS_{amp}$.

Differently from the algorithm by McGehee and McKinney (1995), the algorithm by Leva (2004) monitors the actual propagation of a tsunami by checking the amplitude of the filtered signal against a prescribed threshold ($TS_{amp}$). The algorithm therefore belongs to the class of the *amplitude-discriminating* ones.

Given the duration of the filtering window, such an algorithm has a poor performance in removing meteorological disturbance waves. This algorithm was specifically designed for the TEWS operating at the island of Stromboli. However the tidal wave filtering method by Leva (2004) may also suffice in different but similar situations, depending on the local meteorological wave.

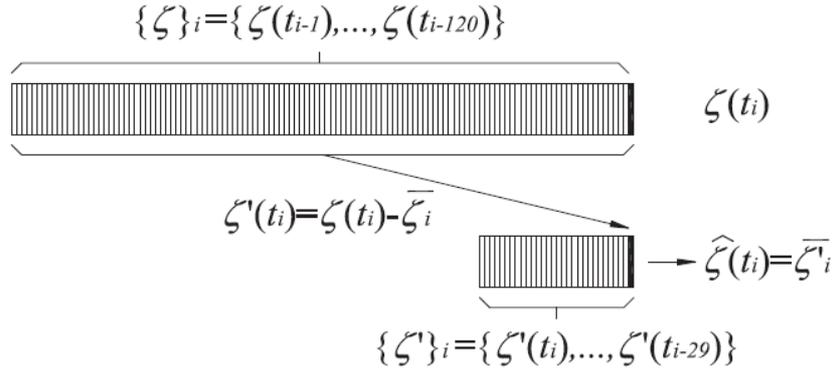

**Figure 3 - Italian Civil Protection filtering method (Leva, 2004)**

### 4.2.4 PARI (Port and Airport Research Institute of Japan) filtering

The filtering algorithm by PARI (Port and Airport Research Institute of Japan) has been developed by Shimizu et al. (2006) to perform the automatic, real-time detection of a tsunami in the sea-level measurements collected by the GPS buoy system (Nagai and Shimizu, 2009) of the Japanese Nationwide Ocean Wave information network for Ports and HArbourS (NOWPHAS).

In order to filter out meteorological wave components, the algorithm by (Shimizu et al., 2006) uses a Finite Impulse Response (FIR) time domain digital filter which relies on the direct use of a Hamming window (Hamming, 1998), which is a window optimized to minimize the nearest side lobe, as impulse response. The filter length is fixed to 120 $s$. Therefore, given the sampling frequency $f_s$, it spans $2M + 1 = 120 \cdot f_s + 1$ samples. Such a length implies that an output delay of 60 $s$ with respect to the actual time should be accepted in order to make the filter physically realizable (causal). Indeed, this is usual for a FIR filter for which half of the impulse response should be applied to samples collected after the filtered one. At the actual time step $t_i$, the output of the filter therefore refers to time step $t_j = t_{i-M}$ ($M = 60 \cdot f_s$), and can be expressed as (Shimizu et al., 2006):

$$\zeta'(t_j) = \zeta'(t_{i-M}) = \sum_{k=-M}^{+M} h_k \cdot \zeta(t_{j-k}) \qquad (XVI)$$

where $\zeta(t)$ is the measured signal, and $\{h\}$ the set of invariant weights known as impulse response. By using a normalized Hamming window as impulse response (Hamming, 1998), the generic weight $h_k$ can be expressed as (Shimizu et al., 2006):

$$h_k = A\left[0.54 + 0.46 \cos\left(\frac{2\pi k}{2M+1}\right)\right] \qquad (XVII)$$

where $-M \leq k \leq +M$, and (Shimizu et al., 2006):

$$A = \frac{1}{\sum_{k=-M}^{+M} 0.54 + 0.46 \cos\left(\frac{2\pi k}{2M+1}\right)}. \qquad (XVIII)$$

is a constant value used to normalize the output signal level (Nagai and Shimizu, 2009).
In order to filter out the astronomical tide, Shimizu et al. (2006) propose to subtract its harmonic prediction $\zeta_p(t_j)$ from the output $\zeta'(t_j)$ of the FIR filter. In particular, it is suggested to carry out such a prediction by using the main four tidal constituents resulting from the harmonic analysis of a 6-day record of previous measurements. Relying on harmonic analysis clearly makes the algorithm site dependent. The prediction at time $t_j$ is expressed as (Shimizu et al., 2006):

$$\zeta_p(t_j) = \sum_c c_c \cdot \cos(\omega_c t_j - \varepsilon_c). \qquad (XIX)$$

where $c_c$, $\varepsilon_c$ and $\omega_c$ are respectively the amplitude, phase and angular frequency of the c$^{th}$ tidal constituent. The filtered sea-level $\hat{\zeta}(t_j)$ at each new time step is finally obtained by carrying out a moving-average analysis of the resulting signal $\zeta''(t) = \zeta'(t) - \zeta_p(t)$. Such analysis is suggested in order to correct the mean sea level filtering out the meteorological tide (Shimizu et al., 2006). As in the case of the Italian Civil Protection filter by Leva (2004), the filter of PARI by Shimizu et al. (2006) belongs to the class of the *amplitude-discriminating* filters. Therefore, it monitors the actual propagation of a tsunami by checking the amplitude of the filtered sea-level against a prescribed threshold ($TS_{amp}$). In other words, a tsunami will be triggered if $|\hat{\zeta}(t_j)|$ exceeds such a threshold.

This FIR filter by Shimizu et al. (2006) is effective in filtering out meteorological waves. However the shorter the period of the tsunami is, the lower the capability of the filtering algorithm of detecting and fully characterizing it. Indeed this filter has been mainly designed for detecting long-period tsunamis (> 10 *min*), and for providing timely warnings of the actually approaching ones not influenced by a delay larger than 1 minute between tsunami measurement and detection (Shimizu et al., 2006).

## 4.3 Infinite-duration Impulse-Response filtering methods

The main disadvantage of Moving-Average filters is that they may require significantly more processing and memory resources than cleverly designed *Infinite-duration Impulse Response* (IIR) filters, although FIR filters are generally easier to design than IIR filters (Mitra and Kaiser, 1993; Hasler and Hasler, 2007).

IIR filters rely on the Butterworth (1930) approximation of the gain or magnitude response function (for more details see (Hasler and Hasler, 2007)). This type of filters has a nonlinear phase response. Therefore, a *bidirectional* filtering of the considered signal segment has to be applied in order to avoid the output phase shift (Shenoi, 2006). The combination of forward and reverse filtering actually produces a zero phase response at the sole cost of doubling the filter execution time.

Considering a time step $t_i$, the output sample y($t_i$) of a *simple* IIR filter is defined by the following so-called "difference equation" (Hasler and Hasler, 2007):

$$y(t_i) = \sum_{h=1}^{q} b_h y(t_{i-h}). \qquad (XX)$$

where $q$ is the order of the filter and $b_h$ are the equation coefficients which may vary for a IIR filter to another. In practice, a simple IIR filter is a "recursive" filter in which the output y($t_i$) is recursively computed from its previously computed values. For this reason a simple IIR filter is also referred to as *Auto-Regressive (AR)* filter (Shenoi, 2006).

The *general* form of an IIR filter is composed of an Auto-Regressive part and a Moving-Average part, and for this reason it is referred to as *ARMA (Auto-Regressive Moving-Average)* filter.
At time step $t_i$, the general form of the output of an ARMA filter can be expressed as (Box and Jenkins, 1976):

$$y(t_i) = \sum_{k=0}^{q} a_k x(t_{i-k}) + \sum_{h=1}^{q} b_h y(t_{i-h}), \qquad (XXI)$$

where $a_k$ and $b_h$ are the weights (or taps) of the filter, $x(t)$ and $y(t)$ respectively the original signal and the filtered signal, and $q$ is the order of the filter. As it can be noticed, the input to the filter consists of samples of both the original and the filtered signal (i.e. of previous filter outputs). It is also to be noticed that, given the sampling rate and the pass and transition band, the filter taps can be calculated once and for all and a priori.
As far as the order $q$ is concerned, its choice is a trade-off between the attenuation rate of the filter and the ringing effects in the filter output due to Gibbs' phenomenon (Gibbs, 1899). The higher the order $q$ the steeper the transition from the stop to the pass band (i.e. the greater the attenuation rate), and the greater the ringing effects (i.e. the presence of a succession of overshoot ripples at both ends of the filtered signal). In the context of real-time tsunami detection this is a major problem since ringing affects the most important part of the filtered signal, i.e. the output at current time.

One of the possible strategies to overcome this limit is to lengthen the chosen segment $\{x\}$ of the actually available (measured) series $x(t)$ with a *fictitious* series $\{x_f\}$ of duration $\Delta t_f$ long enough to avoid, or at least reduce, Gibb's phenomenon at current time $t_i$. The possible method to fulfil this task is that of mirroring $\{x\}$ around current time $t_i$, interposing between $\{x\}$ and its mirror a transition signal $\{x_f\}$ in order to avoid the presence of a discontinuity at $t_i$, that is (Beltrami and Di Risio, 2011; Bressan and Tinti, 2012):

$$x(t_i + t^*) = x(t_i - t^*), \quad \text{with } t^* \in [0, \Delta t_f]. \qquad (XXII)$$

This simple mirroring actually introduces a singular point in the lengthened signal at $t_i$, which can cause problem with detection reliability, depending on the filter cut-off period. A very simple transition signal can be used in the present case. It is based on a sine function and can be expressed as (Bressan and Tinti, 2012):

$$x_t(t) = c_t \sin[\omega_t(t - t_i) + \varepsilon_t], \quad \text{with } t = t_i \ldots t_{i+\gamma}. \quad (XXIII)$$

where $c_t$, $\omega_t$ and $\varepsilon_t$ represent respectively the amplitude, the angular frequency and the phase shift of the transition signal. It is to be noticed that, while the transition signal must satisfy the matching conditions (Beltrami and Di Risio, 2011):

$$x_t(t_i) = x(t_i); \quad \left(\frac{dx_t}{dt}\right)_{t=t_i} = \left(\frac{dx}{dt}\right)_{t=t_i}; \quad (XXIV)$$

its duration (i.e. $t_{i+\gamma} - t_i = \Delta t_t < \Delta t_f$) should be long enough to avoid the introduction of spurious energy in the lengthened signal. Given an angular frequency $\omega_t$, from Equation (XXIII) it is possible to compute $c_t$ and $\varepsilon_t$ as (Bressan and Tinti, 2012):

$$\begin{cases} c_t = \dfrac{\zeta(t_i)}{\cos \varepsilon_t}; \\ \varepsilon_t = \tan^{-1}\left[\dfrac{-(dx_t/dt)_{t=t_i}}{x_t(t_i) \cdot \omega_t}\right]. \end{cases} \quad (XXV)$$

It is clear that the angular frequency $\omega_t$ of the transition signal must be chosen in such a way that it can be filtered out by the IIR filter (Box and Jenkins, 1976).

### 4.3.1 IIR filtering method by Beltrami and Di Risio (2011)

In order to overcome the constraints shown by the FIR algorithms discussed in Section 4.2, Beltrami and Di Risio (2011) propose a method based on a cascade of causal filters, the main one being an IIR time domain filter. The characteristics of the filter by Beltrami and Di Risio (2011) are reported in this section.

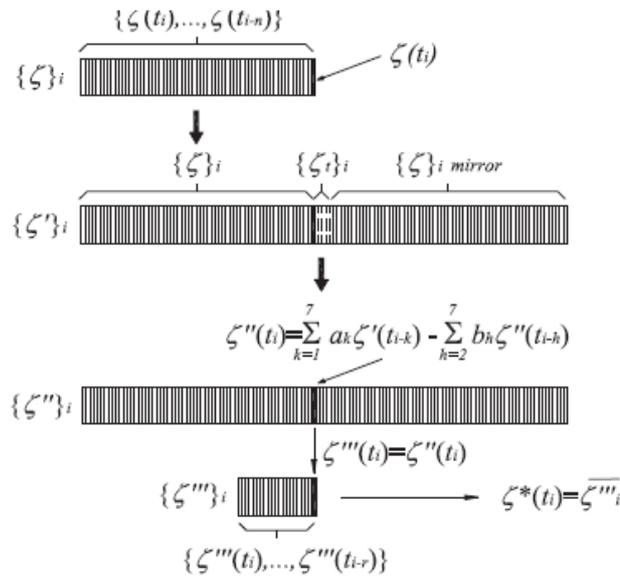

**Figure 4 - Meteorological filtering part of the IIR filter by Beltrami and Di Risio (2011)**

A vector $\{\zeta\}_i = \{\zeta(t_i), ..., \zeta(t_{i-n})\}_i$ consisting of the $n$ samples collected during the time interval $\Delta t_a$ is updated at each time step $t_i$. In order to filter out meteorological waves, the signal stored in $\{\zeta\}_i$ should be previously lengthened by the procedure already illustrated in the preceding section of the fictitious series made of a transition signal $\{\zeta_t\}_i$ and by the mirror of $\{\zeta\}_i$ (see Figure 4). The resulting signal is stored in vector $\{\zeta'\}_i$ of duration $\Delta t = \Delta t_a + \Delta t_f$ (being $\Delta t_f = \Delta t_t + \Delta t_a$).

The following IIR filter of order $q = 7$ (Beltrami and Di Risio, 2011):

$$\zeta''(t_i) = \sum_{k=0}^{7} a_k \zeta'(t_{i-k}) + \sum_{h=1}^{7} b_h \zeta''(t_{i-h}), \qquad (XXVI)$$

is then applied in order to perform the bidirectional filtering of all the samples contained in vector $\{\zeta'\}_i$. The results of the combination of forward and reverse filtering are stored in vector $\{\zeta''\}_i$. The filtered sample at time $t_i$, is then extracted from $\{\zeta''\}_i$ and stored in a further vector $\{\zeta'''\}_i = \{\zeta''(t_i), ..., \zeta''(t_{i-r})\}_i$ consisting of the $(r-1)$ values obtained in the last $r/f_s$ seconds. At each new time step $t_i$, the sea-level sample $\zeta^*(t_i)$ filtered for the "disturbance" due to meteorological waves is finally obtained by taking the mean $\overline{\zeta'''}_i$ of $\{\zeta'''\}_i$.

**Table 1 – Parameters values suggested in the case of a sampling frequency (Beltrami and Di Risio, 2011)**

| Parameter | Value | Units |
|---|---|---|
| $\Delta t_a$ | 30 | Minutes |
| $\Delta t_t$ | 6 | s |
| $r$ | 45 | Samples |

The application of the described part of the filtering algorithm by Beltrami and Di Risio (2011) implies a choice for the values of parameters such as the duration of time intervals $\Delta t_a$ and $\Delta t_t$, and the number $r$ of samples stored in vector $\{\zeta'''\}_i$. This choice clearly depends on the range of periods of the tsunamis to be detected. In
Table **1** it is reported an example for parameters values chosen in order to give to the filtering algorithm the wider range of application, i.e. in order to make it capable of detecting and fully characterizing tsunamis with periods larger than 60 *s*.

As far as the residual tidal (astronomical and meteorological) "disturbance" is concerned, the algorithm proceeds as follows. Each new sample $\zeta^*(t_i)$ is stored in a vector $\{\zeta^*\}_i$ consisting of the samples filtered for meteorological waves during the last 3 *h* and 40 *min*. A specific and varied version of the same cubic polynomial on which the tsunami detection algorithm developed by Mofjeld (1997) relies is then used in order to filter out the tidal wave pattern from the signal $\{\zeta^*\}_i$.

The polynomial is fitted to *p*-minute averages $\overline{\zeta^*}$ (centred at the *p/2 min*) of samples $\zeta^*$ stored over the 3 *h* which precedes of 30 *min* the current time (see Figure 5). Given the sampling frequency $f_s$, the output of the polynomial can be expressed as (Beltrami and Di Risio, 2011):

$$\zeta_p(t_i) = \sum_{j=0}^{3} \omega_k \overline{\zeta^*}\left(t_{i-1800 \cdot f_{s-1}} - 30p - 3600j\right) \qquad (XXVII)$$

where the coefficients are calculated a priori by applying the Newton's forward divided difference formula (Mofjeld, 1997; Beltrami, 2008). Table 2 shows the polynomial coefficients for the case *p* = 10 *min* and $f_s$ = 2 *Hz*. It is to be noticed that these coefficients are not site dependent; the filtering algorithm therefore does not depend on the location of the measurement device.

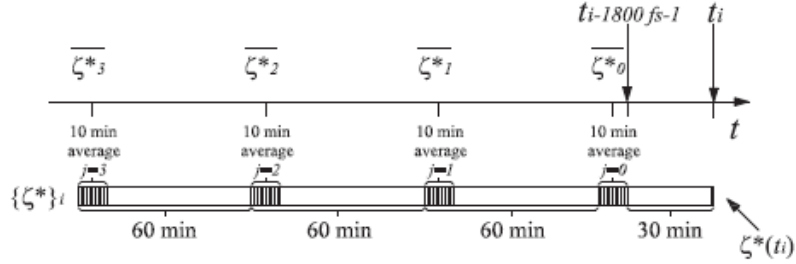

**Figure 5 - Scheme of the cubic polynomial for 10-min averages of samples ζ (Beltrami and Di Risio, 2011)**

**Table 2 - Polynomial coefficients for 10-min averages of observations, p = 10 min and $f_s$ = 2 Hz (Beltrami and Di Risio, 2011)**

| p (minutes) | 10 |
|---|---|
| w0 | +2.4432451059353566 |
| w1 | −2.7008348451980630 |
| w2 | +1.6554065948122720 |
| w3 | −0.3978168555495660 |

The tidal-wave pattern is filtered out by subtracting at each new time step $t_i$ the polynomial output $\zeta_p(t_i)$ from the corresponding sample $\zeta^*(t_i)$. The filtered sea-level at each new time step $t_i$ is therefore expressed as $\hat{\zeta}(t_i) = \zeta^*(t_i) - \zeta_p(t_i)$. As in the case of the FIR algorithms of the Italian Civil Protection filtering by Leva (2004) and of the Port and Airport Research Institute of Japan by Shimizu et al. (2006), see Section 4.2.3 and Section 4.2.4, a tsunami will be triggered if $|\hat{\zeta}(t_i)|$ exceeds a preselected threshold $TS_{amp}$ (Beltrami and Di Risio, 2011).

A possible alternative to the cubic polynomial by Mofjeld (1997) is the use of artificial neural networks. (ANNs). How it will be described successively in the paper, a well-designed ANN can guarantee generally a better filtering for tidal "disturbance". However, its implementation makes the overall algorithm site dependent (Beltrami, 2008).

## 4.4   Kalman filtering

In contrast to the simplicity of its physical model, the Kalman filter (Kalman, 1960), widely used in engineering practice, is by some criteria a more sophisticated data assimilation procedure than any now in operational use in meteorology or oceanography. If certain conditions are met, it can be shown to yield an estimate of the present state of a system, which is statistically optimal. In some respects, the sea level forecasting problem is an ideal candidate for Kalman filtering. The evolution equations are predominantly linear. The errors result largely from errors in the forcing and, in some circumstances, from model deficiencies, as opposed to the loss of predictability inherent in nonlinear dynamics. In the present section it is described how the Kalman filter is used to analyse and forecast the sea level height. Many different variants are possible and present in the literature, e.g. (Altaf et al., 2013; Choi et al., 2000; Sørensen et al., 2006; Yen et al., 1996), but the very general form is that described in the following.

We assume a model of the sea level height represented by a series of vectors $\alpha_t$, also referred to as *state vectors*. These variables are supposed to describe the current state of the sea level system to be analysed. These state variables will typically not be observed and the other main ingredient is therefore the observed variables of the sea level, $y_t$. The first step consists of expressing the model in state space form that is in the form of a linear system formed by two sets of linear equations. The first set of equations describes the evolution of the system and is called the *Transition Equation* (Miller, 1986):

$$\alpha_t = K\alpha_{t-1} + R\eta_t, \qquad (XXVIII)$$

where $K$ and $R$ are matrices of constants and $\eta$ is $N(0,Q)$. The second set of equations describes the relation between the state of the system and the sea level observations, and it is called the *Measurement Equation* (Miller, 1986):

$$y_t = Z\alpha_t + \varepsilon_t, \qquad (XIX)$$

where $\varepsilon_t$ is $N(0,H)$ and $E\left(\varepsilon_t \eta_{t-j}\right) = 0, \forall t, j$.

Sea level models can be put in the state space form under some assumptions; the main restriction is of course on the linearity of the model. The state-space model as it is defined here is not the most general possible. It is in principle easy to allow for non-stationary coefficient matrices, see for example Harvey (1989). There exist also extensions of Kalman filter methods to non-linear models. These are known as *extended Kalman filters* and are not necessarily able to evaluate likelihood functions exactly but only produce approximations (for more details see (Altaf et al., 2013; Yen et al., 1996)).

As showed in Section 4.3, sea level systems can be expressed as Auto-Regressive Moving-Average (ARMA) processes. An ARMA process can also be written in State-Space form; thus the Kalman filter can be used to estimate its likelihood function. In the case of a general ARMA process it is possible to use several representations (see Section 4.3).

In this context assume to have a scalar ARMA process (Harvey, 1989):

$$x_t = a_1 X_{t-1} + \ldots + a_k X_{t-m} + u_t + b_1 u_{t-1} + \ldots + b_l u_{t-l}, \qquad (XX)$$

where $m = max\{k, l+1\}$. A scalar ARMA is one of the most compact and easy forms for a general ARMA system. Here the mean value is omitted to simplify the exposition. This process can be represented in State-Space form by the following transition equation (Harvey, 1989):

$$\alpha_t = \begin{pmatrix} a_1 & 1 & 0 & \cdots & 0 \\ a_2 & 0 & 1 & \cdots & 0 \\ \vdots & \vdots & \vdots & \ddots & \vdots \\ a_{m-1} & 0 & 0 & \cdots & 1 \\ a_m & 0 & 0 & \cdots & 0 \end{pmatrix} \alpha_{t-1} + \begin{pmatrix} 1 \\ b_1 \\ \vdots \\ b_{m-2} \\ b_{m-1} \end{pmatrix} u_t \qquad (XXI)$$

and measurement equation:

$$x_t = (1, \ 0, \ \cdots, \ 0) \, \alpha_t. \qquad (XXII)$$

The Kalman filter is very useful in order to calculate a likelihood function. Recursive likelihood equations, like the equations in a sea-level model, allow evaluating the "impact" of a new observation arriving, in the sense that it immediately shows the conditional likelihood. The Kalman Filter is able to update a parameter estimate instantly when a new observation occurs, without having to re-estimate using all the data. Use of the Kalman filter is a particular good choice for the purpose of single estimations on a given data set, as in tide analysis and forecasting equations (Choi et al., 2000).

In order to evaluate the value of a likelihood function for a given set of parameters it is typically necessary to use a general maximization algorithm whose parameters are updated by a Kalman filtering algorithm taken as input subroutine for the maximization algorithm (Yen et al., 1996).

Consider a general likelihood function (Harvey, 1989):

$$f(y_T, \ \cdots, \ y_1, \ \theta), \qquad (XXIV)$$

where $\theta$ is a vector of parameters. Independently by the physical model or distribution that generates the variables, it is always possible to factor such a likelihood function as:

$$f(y_T, \ \cdots, \ y_1, \ \theta) = f(y_T | y_{T-1}, \ \cdots, \ y_1, \ \theta) \cdot f(y_{T-1}, \ \cdots, \ y_1, \ \theta). \qquad (XXV)$$

By iterating it is obtained the following formula:

$$f(y_T, \ \cdots, \ y_1, \ \theta) = \prod_{t=p+1}^{T} f(y_T | y_{T-1}, \ \cdots, \ y_1, \ \theta) \cdot f(y_p, \ \cdots, \ y_1, \ \theta), \qquad (XXVI)$$

yielding to the following log-likelihood function in *recursive form* (Harvey, 1989):

$$L = \ln f(y_T, \ \cdots, \ y_1, \ \theta) = \sum_{t=p+1}^{T} f(y_T | y_{T-1}, \ \cdots, \ y_1, \ \theta) + f(y_p, \ \cdots, \ y_1, \ \theta). \qquad (XXVII)$$

In the case of the normal likelihood function this becomes:

$$L = \sum_{t=p+1}^{T} -\frac{1}{2}|F_t| - \frac{1}{2}v'_t F_t^{-1} v_t + constant + \ln f(y_p, \ \cdots, \ y_1, \ \theta), \qquad (XXVIII)$$

where (Harvey, 1989):

$$\begin{cases} v_t = y_t - E(y_t|y_{t-1}, \cdots, y_1); \\ F_t = E(v_t v'_t|y_{t-1}, \cdots, y_1). \end{cases} \quad (XXIX)$$

The Kalman filter equations, as it will be shown in the following, are the ones used to evaluate $F_t$ and $v_t$, which is a quite convenient way of evaluating the likelihood function, so allowing the prediction and the update of the model equations.

For any vector $x_t$ define $x_{t|t-1} = E(y_t|y_{t-1}, \cdots, y_1)$, where $y_j$ are the observed variables. This definition gives the best guess of $x_t$ based on all the information available at time $t-1$, while $x_{t|t-1}$ is the prediction of $x_t$ at $t-1$. It is intuitive that the Kalman filter develops around predicting and updating the prediction of the state vector. Also define $P_{t|t-1} = E\{(\alpha_t - \alpha_{t|t-1})(\alpha_t - \alpha_{t|t-1})'\}$ where $P_{t|t-1}$ is the conditional variance of the "prediction error" $\alpha_t - \alpha_{t|t-1}$.

In the following it is first described the Kalman filter and then how it is derived.
The prediction equations take the following form (Altaf et al., 2013):

$$\begin{cases} \alpha_{t|t-1} = K \cdot \alpha_{t-1|t-1}; \\ y_{t-1} = Z \cdot \alpha_{t|t-1}, \\ P_{t|t-1} = K \cdot P_{t-1|t-1} \cdot K' + R \cdot Q \cdot R' \end{cases} \quad (XXX)$$

Now define the "variance matrix" as (Harvey, 1989):

$$v_t = y_t - y_{t|t-1}, \quad (XXXI)$$

Since $y_t - y_{t|t-1} = Z(\alpha_t - \alpha_{t|t-1}) + \xi_t$ it follows that:

$$F_t = E\{v_t v'_t\} = E\{(y_t - y_{t|t-1})(y_t - y_{t|t-1})'\} = Z \cdot P_{t|t-1} \cdot Z' + H. \quad (XXXII)$$

To complete the Kalman filter, the updating equations are required (Yen et al., 1996):

$$\begin{cases} \alpha_{t|t} = \alpha_{t|t-1} + P_{t|t-1} \cdot Z' \cdot F_t^{-1} \cdot v_t; \\ P_{t|t} = P_{t|t-1} - P_{t|t-1} \cdot Z' \cdot F_t^{-1} \cdot Z \cdot P_{t|t-1}. \end{cases} \quad (XXXIII)$$

The interpretation of the updating equations is that $v_t$ contains the new information (from $y_t$) and the estimate of $\alpha_t$ based on $y_1, \cdots, y_{t-1}$ (i.e. $\alpha_{t|t-1}$), is updated to a new estimate that is based on $y_1, \cdots, y_{t-1}$ and $y_t$ (i.e. the new estimate is $\alpha_{t|t}$), and the conditional variance $P_{t|t}$ of $\alpha_{t|t}$ is then calculated. The term $P_{t|t-1} \cdot Z' \cdot F_t^{-1} \cdot v_t$ is called "Kalman gain", and any new information enters the system through this term.

The Kalman filter can be derived from the rules of the Normal distribution. It is possible to write (Harvey, 1989):

$$\begin{cases} v_t = Z(\alpha_t - \alpha_{t|t-1}) + \xi_t; \\ \alpha_t = \alpha_{t|t-1} + (\alpha_t - \alpha_{t|t-1}). \end{cases} \quad (XXXIV)$$

From this definition it is obtained:

$$\begin{pmatrix} v_t \\ \alpha_t \end{pmatrix} = N\left( \begin{pmatrix} 0 \\ \alpha_{t|t-1} \end{pmatrix}, \begin{bmatrix} F_t & Z \cdot P_{t|t-1} \\ P_{t|t-1} \cdot Z' & P_{t|t-1} \end{bmatrix} \right) \quad (XXXV)$$

Now in the following it is recalled a special rule for the conditional Normal distribution (see (Lütkepohl, 1993)) for more details). Standing the following statement:

$$\begin{pmatrix} x_1 \\ x_2 \end{pmatrix} \approx N\left( \begin{pmatrix} \mu_1 \\ \mu_2 \end{pmatrix}, \begin{bmatrix} \Sigma_{11} & \Sigma_{12} \\ \Sigma_{21} & \Sigma_{22} \end{bmatrix} \right), \quad (XXXVI)$$

then the conditional distribution of $x_1$ given $x_2$ is (Lütkepohl, 1993):

$$N(\mu_1 + \Sigma_{12}\Sigma_{22}^{-1}(x_2 - \mu_2), \quad [\Sigma_{11} - \Sigma_{12}\Sigma_{22}^{-1}\Sigma_{21}]). \quad (XXXVII)$$

Using this rule on the conditional distribution of $(v_t, \alpha_t)$ it is obtained (Harvey, 1989):

$$f(\alpha_t|y_t, \ldots, y_1) = f(\alpha_t|v_t, y_{t-1}, \ldots, y_1) = \cdots$$
$$\ldots = N(\alpha_{t|t-1} + P_{t|t-1} \cdot Z' \cdot F_t \cdot v_t, \quad P_{t|t-1} - P_{t|t-1} \cdot Z' \cdot F_t^{-1} \cdot Z \cdot P_{t|t-1}), \quad (XXXVIII)$$

which is finally the updating equation.

One of the main problems is how to initialize the filter. It is natural to choose $\alpha_{t|t-1} = 0$ since this is the unconditional mean of $\alpha_t$. In addition it is also natural to choose the stationary value of the variance as the initial value, even though this is only an option in the stable case. Other choices are possible. For example in some circumstances it is required to condition on initial values (as discussed earlier), but in this case the special conditions have to be considered one by one. However it is also possible to find the stationary variance by the following method.

Combining the updating and the prediction equations it is obtained:

$$P_{t+1|t} = K \cdot P_{t|t-1} \cdot K' - K \cdot P_{t|t-1} \cdot Z'(Z \cdot P_{t|t-1} \cdot Z' + H)^{-1} Z \cdot P_{t|t-1} \cdot K' + R \cdot Q \cdot R', \quad (XXXIX)$$

which is known as the *Riccati equation* (Zwillinger, 1997). If the model is stable $P_{t|t-1}$ will converge to the solution $\bar{P}$ of the algebraic Riccati equation (Zwillinger, 1997):

$$\bar{P} = K \cdot \bar{P} \cdot K' - K \cdot \bar{P} \cdot Z'(Z \cdot \bar{P} \cdot Z' + H)^{-1} Z \cdot \bar{P} \cdot K' + R \cdot Q \cdot R'. \quad (XL)$$

As already said, the most natural choice for the starting values of a stable system is to choose unconditional mean and variance. Since $\alpha_t = K \cdot \alpha_{t-1} + R \cdot \eta_t$ has the form of an AR(1) model (see Section 4.3), it is chosen $\alpha_{1|0} = 0$ and $P_0$ such that (Harvey, 1989):

$$vec(P_0) = (I - K \otimes K)^{-1} vec(R \cdot Q \cdot R'). \qquad (XLI)$$

It is also possible to choose other initial conditions, for example from a Bayesian prior, to condition on initial values. However, in the non-stationary case like a sea-level distribution, obviously it is not possible to choose the initial distribution from the stationary distribution. To conclude, in order to use the Kalman filter it is necessary to dispose of a general optimization routine from the literature in order to minimize (or maximize) the value of the likelihood function.

As already stated in the literature review section (Section 2), a large application of Kalman filters to sea-level models is found in the literature. In particular the Kalman filter by Miller and Cane (1989) for a simple model of sea level anomalies in the tropical Pacific was using tide gauge data from six selected island stations to update the model and required detailed statistical assumptions about the errors in the model and the data. The error model in their filter was a simple covariance function with parameters fit from the observed differences between the tide gauge data and the model output. The calibrated error model was used in the Kalman filter to generate monthly sea level height anomaly maps. The filtered maps exhibited fine structure that is absent from the unfiltered model output, even in regions removed from the data insertion points. Given that only six data points were embedded in the data assimilation, the results obtained by the Kalman filter of Miller and Cane (1989) were quite encouraging, and they were further refined by the Kalman filtering method for the tropical Pacific in (Chan et al., 1996). In the Kalman filtering method proposed by Yen et al. (1996), a harmonic tide level model was used to predict tide levels of the Kaohsiung Harbor in Taiwan. The parameters of the tide level model (i.e. the amplitudes of the harmonic components) were estimated by Kalman filtering technique using just a few-day tide record with the assumption of known angular frequencies. Their predicted tide levels were in good agreement with the observed real data. Other tide analysis and forecasting methods based on Kalman filtering were proposed successively in the literature. The most relevant perhaps is the contribution by Choi et al. (2000), in which sea-level predictions from short observation record, based on harmonic tidal models with their parameters estimated by the Kalman filter algorithm, were tested for Macau. Afterwards Sørensen et al. (2006) presented a parameter sensitivity study of three well known general Kalman filter approaches (the ensemble Kalman filter, the reduced rank square root Kalman filter, and the steady Kalman filter) to use for the assimilation of water levels in a three dimensional hydrodynamic modelling system. Their sensitivity study further demonstrates the effectiveness of Kalman filtering in dealing with this kind of systems. In addition Altaf et al. (2013) has strengthen the ensemble Kalman filter for storm surge determining harmonic parameters with a limited amount of tidal measured data by robust adaptive inflation.

# 5 Wavelet transform analysis

Tidal analysis shows some particular differences from most other geophysical time-series analysis. The most important is the marked contrast between a "short-term tidal signal" (i.e. a daily record), where even non-stationary flows exhibit a strong dominance by processes occurring within narrow frequency ranges known from astronomical considerations, and the "long-term tidal signals" (i.e. monthly or yearly records), where stochastic forcing with a highly variable and broad-band frequency structure is usually seen. Thus in tidal analysis is necessary to use information concerning astronomical forcing without introducing assumptions that obscure non-tidal processes in a record, and to provide an internally consistent extraction of both tidal and non-tidal variance. Furthermore, in contrast to many areas of geophysics where analysis of frequency content is the sole objective, in tide analysis it is often requested to produce also forecasts of tide sea level as well as to reconstruct the original data whenever observations are missing.

The requirements of dynamical analysis and the large volume of data to analyze require a methodology that is able to analyze all variables and all frequencies of the tide signal in a consistent manner. This section describes a tidal analysis methodology based on continuous wavelet transforms (CWTs) first proposed by Flinchem and Jay (2000) and that is able to provide a consistent, linear analysis of tidal and non-tidal variance and reveals features that harmonic analysis is not able to elucidate.

The simpler forms of the analysis/prediction problem have been solved primarily by harmonic analysis methods, but many significant quasi-periodic tidal processes less amenable to study and to predict still remain opened. Tidal phenomena may be irregular either because an aperiodic input is competing with tidal forcing or the oceanic response to tidal forcing is being modulated by some internal process. Variable wind stress and sea-level pressure are examples of the former. Practical forecasting of such phenomena would require both an invertible analysis method suitable to non-stationary processes, and a method of forecasting the stochastic processes that modify the astronomical tide. The practicality of the second task depends on the problem; thus CWT analysis concentrates on accomplishing the first task (Labat, 2005).

The forecasting power of harmonic analysis derives from its representation of the tide by a line spectrum consisting of a finite number of infinitely narrow peaks at fixed, predetermined frequencies, each with a definite amplitude and phase. A time series with such a spectrum is a stationary signal, i.e. one that may be divided into multiple, statistically indistinguishable segments. Of course reality is not so simple, and the presence of noise must be acknowledged even for stationary tidal processes.

A non-stationary signal has, in contrast, a frequency content that evolves over time. If a non-stationary times series is dissected, the statistical properties of the parts will not all be similar to each other or to the statistics of the whole, and the usual Gaussian statistics based on the whole record will be deceptive (Sun et al., 2006; GÓmez-Cubillo et al., 2012).

In oceanography, non-stationary tidal signals provide the opportunity to deepen our understanding of tidal dynamics. Harmonic analysis converts the information content in the time domain of a signal into static, averaged frequency information. This limitation of harmonic analysis methods to yield only a static picture of the frequency content of a non-stationary signal is overcome by wavelet techniques which, as it will be shown in the following, are able to translate the idea of evolving frequency content into mathematics, turning intuitions about non-stationary signals into a useful scientific tool.

To define the continuous wavelet transform (Flinchem and Jay, 2000), consider an oscillatory prototype function $\Psi_0(t)$ with zero mean, finite variance and that is localized in time near the origin, i.e.:

$$\int \Psi_0(t)\, dt = 0, \quad \int \Psi_0(t) \cdot \overline{\Psi_0}(t) > \infty, \quad \lim_{|t|\to\infty} \Psi_0(t) = 0, \qquad (XLII)$$

where integration is over the entire real line. These constraints are quite broad, and there is wide latitude in the choice of the form of $\Psi_0(t)$. These properties guarantee that the wavelet is "wavelike" (has no zero-frequency energy) and localized in time-frequency space (Farge, 1992). From $\Psi_0(t)$ a two-parameter family of functions is defined by scaling and translating the argument (Figure 6). For $0 > a > \infty$ and $-\infty > b > \infty$:

$$\Psi_{a,b}(t) = a^{-p}\, \Psi_0\left(\frac{t-b}{a}\right). \qquad (XLIII)$$

If $a$ and $b$ are continuous, $\Psi_{a,b}(t)$ forms a complete basis for $L^2(\varphi)$, analogous to the Fourier integral transform basis set $[\exp(i2\pi \Psi t)]$ over t $[-\infty, +\infty]$. Setting $p = 0.5$ ensures that all basis functions have the same variance, regardless of scale, i.e.:

$$\int |\Psi_0(t)|^2\, dt = \int |\Psi_{a,b}(t)|^2\, dt, \quad \forall a,b,c > 0. \quad (XLIV)$$

Choosing $p = 1$ causes an input wave with unit amplitude 1 to have unit output for all scales.

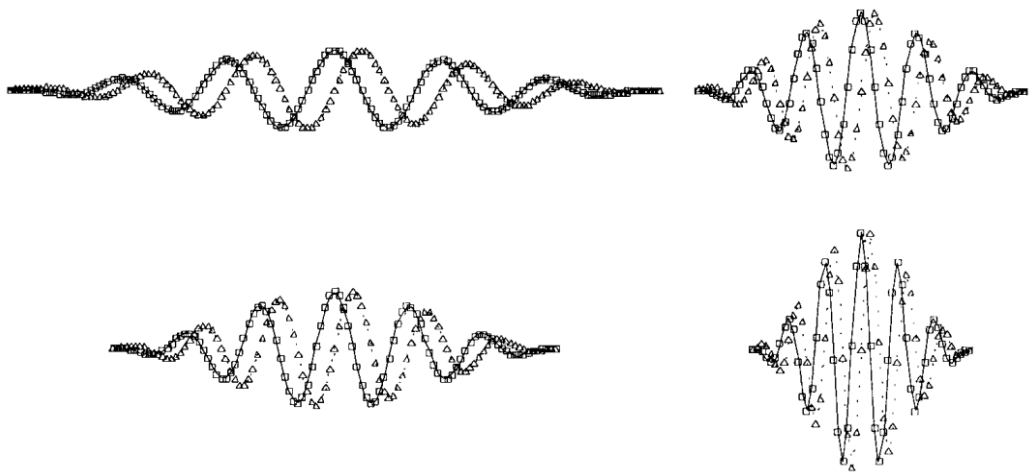

**Figure 6 - Real (□) and imaginary (Δ) parts of typical Kaiser-filter D1, Dint, D2 and D3 wavelets used for hourly data. Each filter has c. 6 cycles; the D1 filter is 145 hours long. The scaling is such that a unit input wave at each scale yields unit output amplitude (Flinchem and Jay, 2000).**

The forward wavelet transform is a convolution similar to a Fourier transform (Farge, 1992):

$$\check{Z}_{a,b=g_{a,b}}[Z(t)] \equiv \int Z(t)\, \Psi^*_{a,b}\, dt, \qquad (XLV)$$

where the inverted hat denotes the transformed quantity) so that $\check{Z}$ is the CWT of $Z$. Like the Fourier transform, the CWT has an inverse or synthetic form (GÓmez-Cubillo et al., 2012):

$$\begin{cases} Z(t) = C_0 \iint \check{Z}_{a,b}(t)\, \Psi_{a,b}\, da\, db, \\ \\ C_0 = \left(2\pi \int \frac{|\widehat{\Psi}_0|^2}{\omega}\, d\omega\right)^{-1}. \end{cases} \qquad (XLVI)$$

The completeness of the set $\{\Psi_{a,b}(t)\}$ means that Equation (XLV) and Equation (XLVI) form a reversible transform pair, analogous to Equation (XLII). Consequently, the CWT shares with the Fourier transform the desirable property of conserving variance. The inverse formula, Equation (XLVI), valid in continuous time and scale domain (scale is inverse to frequency), has an analogue in discrete, finite implementations, but a bondable error is then incurred (Kaiser, 2011).

Wavelet transforms are able to convert a one-dimensional input signal (a function of time) into a two-dimensional field showing the amplitude and phase behavior of the input as a function of both frequency and time. A key link is the Heisenberg uncertainty relation (Gabor, 1946):

$$\sigma_\omega\, \sigma_t \geq \frac{1}{4\pi} \qquad (XLVII)$$

which says that the product of the uncertainty in time (i.e. the duration), $\sigma_t$, and the uncertainty in frequency, $\sigma_\omega$, is greater than (or equal to) a constant value. These terms have meanings analogous to the concept of the standard deviation in statistics. The meaning of this equation is that, for any method of time-frequency analysis, the possibility of knowing simultaneously and mathematically the time evolution and frequency content of data is well limited (Gabor, 1946). However this equation is not an empirical law; rather, it follows from the definition of frequency (or wavenumber) itself. At issue is not noise or imperfect calculations, but simply that it is inconsistent for the two questions: "when?" and "what frequency?", both to have exact answers with respect to the same data. Infinite precision in frequency implies infinite duration in time. On the other hand, exact time precision precludes any knowledge of frequency.

Harmonic analysis of a long, stationary record represents a limiting case of the Heisenberg principle in which the duration of the record, $\sigma_t$, is large and $\sigma_\omega$ is accordingly very small, allowing resolution of closely spaced frequencies. To examine evolution of frequency content in a non-stationary record, it is necessary to choose a smaller variable $\sigma_t$, with the consequence that $\sigma_\omega$ will grow in inverse proportion (Flinchem and Jay, 2000).

A continue wavelet transform approach is complementary to harmonic analysis with respect to the Heisenberg principle, which define the minimum product of time and frequency uncertainties. It is clear from Equations (XIII) and (XLV) that (Flinchem and Jay, 2000):

$$\sigma_t(\Psi_{a,b}) = a\,\sigma_t(\Psi_0), \quad \sigma_\omega(\Psi_{a,b}) = a^{-1}\,\sigma_\omega(\Psi_0). \quad (XLVIII)$$

Therefore, the CWT maintains the Heisenberg relation $\sigma_\omega\,\sigma_t = costant$, with $costant$ close to the optimum allowed by Equation (XLVII) for all basis vectors across all scales. This is a property unique to wavelet transforms, and one of the key facts that make them useful.

Another essential property of wavelet transforms is their linearity; they are indeed additive and distributive (see (Holschneider, 1995; Labat, 2005) for more details). All this properties guarantee that wavelets' results in one frequency band are independent of those in other bands, so that the frequency responses of a wavelet and of a CWT analysis using a series of wavelet filters are well-defined functions. This is emphatically not the case for harmonic analysis, and short harmonic analysis windows have frequency responses that depend on the details of the data analyzed and the number of analysis frequencies chosen (GÓmez-Cubillo et al., 2012).

## 5.1   Application of continuous wavelet transforms to tidal analysis

In order to apply wavelets to tidal analysis, it is necessary to discretize variables $a$ and $b$ in Equations (XLIII) and (XLV), with $a$ chosen to match tidal frequencies. Certain choices of $a \leq 2$, $b$ and $\Psi_0$ will still form a complete basis. In most wavelet applications, data compression is optimized and redundancy avoided by expressing $a$ and $b$ in geometric series, so that the time step $b_n$ increases in size along with scale $a_n$ (Zhong and Oyadiji, 2013):

$$a_n = a_0^{\,n}, \quad 1 < a_0 < 2, \quad b_n = m \cdot b_0^{\,n}, \quad \text{for } n, m = 0, 1, 2, \ldots \quad (XLIX)$$

As the step sizes are increased, there is a limit at which a complete representation becomes impossible. Only for very restricted choices of $\Psi_0$ it is possible to set $a_0 = b_0 = 2$ and still have a complete, orthogonal discrete wavelet transform basis (Flinchem and Jay, 2000). Redundancy, in the form of several filters per octave, eases the difficulty in conserving variance and allows for a wider range of wavelet choices (see (Daubechies, 1988) or (Kaiser, 2011) for further details). To define a wavelet basis suitable for tidal problems, it is necessary to sacrifice data compression and employ a continue wavelet transform approach that (Flinchem and Jay, 2000):
  a) uses wavelets that resemble physical waves;
  b) has frequencies selected according to the dictates of astronomy;
  c) retains the property of completeness, so that the transform may be inverted.
In practice, this requires non-integral $n$.

The very broad bandwidth of tidal signals also affects data compression and invertibility. Most tidal records are short enough relative to the periods of the signals involved so that data compression possibilities are limited, and they fail to capture even one cycle of strong low frequency oscillations present in the tidal signal. The result is that no transformation of a tidal record, whether using harmonic analysis or wavelet transform, can perfectly capture the low-frequency energy, resulting in an imperfect representation of the data after inversion (Flinchem and Jay, 2000; GÓmez-Cubillo et al., 2012).

However, despite its imperfect analysis of low-frequency variance, continue wavelet transform methods still possess an important advantage for tidal records: frequency spacing can be made more or less constant with scale, rather than decreasing at the low-frequency end (as with harmonic analysis).

A dense, even sampling in scale and time yields only small errors in reproduction of the data after transformation; sparse sampling introduces larger errors. A partially redundant approach with more than one filter per octave is useful for tidal analysis, because this (Flinchem and Jay, 2000):
  a) matches the expected tidal frequencies;
  b) eases restrictions on wavelet form;
  c) provides robustness in the face of noisy data;
  d) allows a smooth transition between the distinct analysis needs in the tidal and sub-tidal bands.

The result is a "snug frame" that approximately conserves variance and allows reconstruction of the signal from its continuous wavelet transform.

The idea that analysis needs for the tidal and subtidal bands require rather different approaches bears further explanation. In typical wavelets applications (e.g. image processing) there are no preferred frequencies, and the emphasis is on representation of the data as completely and compactly as possible. But even non-stationary tidal processes usually show, within the tidal band, a concentration of energy at tidal frequencies. Considering this and other factors, most investigations of tidal dynamics should (Zhong and Oyadiji, 2013):
  a) employ wavelets that look as much like a linear wave as possible;
  b) provide information concerning processes occurring at tidal frequencies,
  c) allow inversion (and thus, possible prediction),
  d) minimize overlap between tidal frequencies, without presuming that non-tidal processes are absent in the tidal band.

The spacing of major tidal species $D_1$ (diurnal), $D_2$ (semidiurnal), $D_3$ (terdiurnal), $D_4$ (quarterdiurnal), $D_6$ (sixth-diurnal), $D_8$ (eighthdiurnal), …, up to the Nyquist frequency gives two filters per octave, except between $D_1$ and $D_2$. This series is completed by adding filters known as (Flinchem and Jay, 2000):
  a) $D_{Int}$, centred at the local inertial frequency, conveniently between D1 and D2 in temperate latitudes;
  b) $D_H$ centred at $(1 \cdot 41 \text{ day})^{-1}$, to match smoothly to subtidal filters.

This scheme gives $a_0 = \sqrt{2}$ in Equation (XLIX), with all values of *n* close to integral. The value of $\Psi_0$ is chosen so that overlap between the tidal species is as small as possible. The result is a good compromise (Flinchem and Jay, 2000): there is some unevenness in frequency spacing to accommodate tidal and inertial signals, but (because of finite filter width) little loss of non-tidal variance present in the tidal band. One may further add species such as $D_5$, $D_7$, etc. …, if desired, though this will increase overlap between frequencies.

There are, in contrast to the tidal band, usually no preferred frequencies in the subtidal band. Given the more or less continuous frequency distribution, overlap between frequency bands is desirable, for robustness against noise and completeness. The approach used by Flinchem and Jay (2000) consists of reducing the number of "wiggles" (decrease) in the basis function $\Psi_0$ for the subtidal filters, shortening the subtidal filters while broadening their frequency response. It is then possible to represent subtidal frequencies with one filter per octave (i.e. $a_0 = 2$ in Equation (XLIX)) and simultaneously to reach slightly lower frequencies.

It is also necessary to choose a specific wavelet $\Psi_0(t)$ to implement. A major factor in this regard is the problem of energy leakage into the side-lobes. Optimal for minimizing side-lobe energy under discrete convolution are the prolate spheroidal wave functions (Zhong and Oyadiji, 2013). Unfortunately, these functions cannot be expressed in a closed form convenient for computation. However, Kaiser (2011) derived an accurate approximation to the prolate windows in terms of zero-order modified Bessel functions, $I_0(t)$. The Kaiser window also has the highly desirable effect of minimizing the Heisenberg uncertainty relation (Equation (XLVII)) to the point that $\sigma_\omega \sigma_t \approx 1/4\pi$, in a sense achieving the best transform theoretically possible. The wavelet employed by Flinchem and Jay (2000) for tide analysis is constructed by windowing a complex exponential with a Kaiser filter:

$$\begin{cases} \Psi_0(t) = \dfrac{I_0\left(\beta\sqrt{1-t^2}\right)}{I_0(\beta)} e^{i2\eta\pi t} |t| \leq 1, \quad \text{where } \Psi_0(t) = 0 \text{ if } |t| \geq 1, \\ \Psi_{a,b}(t) = \dfrac{1}{a} \cdot norm(\eta, a) \cdot \Psi_0\left(\dfrac{t-b}{a}\right), \end{cases} \quad (L)$$

where: side-lobe suppression is controlled by $\beta$, the number of cycles on either side of the central point is $\eta$, and $norm(\eta, a)$ is a normalization factor that provides unit response for waves of unit amplitude. If frequencies are evenly spaced, $norm$ is independent from $a$.

Filter design always involves a balancing of factors, such as side-band rejection (i.e. frequency resolution) and filter length (i.e. time resolution). For the Kaiser filter for tidal analysis specified above, the proper parameters setting derived by the authors is available in (Flinchem and Jay, 2000). Similar results to those obtained by this Kaiser filter could have been achieved by using, for example, a Morlet function or other complex wavelet functions (Labat, 2005; Sun et al., 2006); however the Kaiser filter show to be more efficient in suppressing sidebands for tidal analysis purposes (Flinchem and Jay, 2000; Zhong and Oyadiji, 2013).

# 6   Artificial Neural Networks

The previous sections presented the details of a number of tidal analysis and forecasting methods in the literature. When data for the observed periods are lost or incomplete, methods like harmonic analysis and Kalman filtering are not effective in supplementing the lost data. Therefore, in such cases it is important to find an accurate tidal level prediction technique. For this reason, recently artificial neural networks (ANNs) have been used in the literature as an alternative approach. Based on limited field data, the neural network method can predict hourly, daily, weekly or monthly tidal level more accurately than, for example, harmonic analysis methods.

The origin of neural networks was the bionomics. The biological brain consists of billions of highly interconnected neurons forming a neural network. Human information processing depends on this connectionist system of nervous cells. Based on the advantage of this information processing, neural networks can easily exploit the massively parallel local processing and distribute storage properties in the brain.

The major distinction between the human and the computer in information processing is the ability of pattern recognition and learning. The computer can treat large amounts of data with high speed, but it is not able to solve alone complex tasks such as the classification, optical text recognition, data compression, and learning. On the other hand, human beings can easily recognize the aforementioned works by operating the information processing with highly distributed transformations through thousands of interconnected neurons in the brain (Hirose, 2013).

Generally speaking, ANN is a mathematical system, which can model the ability of biological neural networks by interconnecting many of the simple neurons. The neuron accepts inputs from a single or multiple sources and produces outputs by simple processing with a predetermined non-linear function as shown in Figure 7.

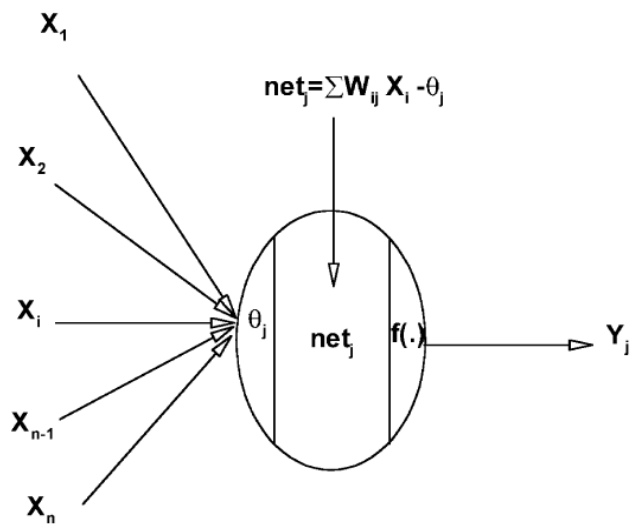

**Figure 7 - An artificial neural symbol**

The training process of the neural network is essentially executed through a series of patterns. In the learning process, the interconnection weights are adjusted within the input and output values. Therefore, the primary characteristics of ANN can be presented as (Rumelhart et al., 1986):
   a) the ability to learn;
   b) distributed memory;
   c) fault tolerance;
   d) parallel operation.

Typically, an ANN is developed as a three-layer learning network. A three-layered network with an input layer (I), a hidden layer (H) and an output layer (O) (see Figure 8) is the most common choice adopted in general. Each layer consists of several neurons and the layers are interconnected by sets of correlation weights. The neurons receive inputs from the initial values or the interconnections. This information will be transferred to the hidden layer, and produce an output with an adequate non-linear *transfer function*. A common transfer function is the sigmoid function expressed by (Hirose, 2013):

$$f(x) = \frac{1}{1+e^{-x}}, \qquad (LI)$$

which has the property:

$$\frac{df}{dx} = f(x)[1-f(x)]. \qquad (LII)$$

Additionally, the learning process (or training) is formed by adjusting the weight of interconnected neurons.
Back-propagation neural networks (BPN), developed by Rumelhart et al. (1986), is the most prevalent of the supervised learning models of ANN. BPN uses the *gradient steepest descent method* to correct the weight of the interconnected neuron. BPN easily solves the interaction of processing elements by adding hidden layers. In the learning process of BPN, the interconnection weights are adjusted using an error convergence technique to obtain a desired output for a given input. This learning process is discussed in detail in the following.

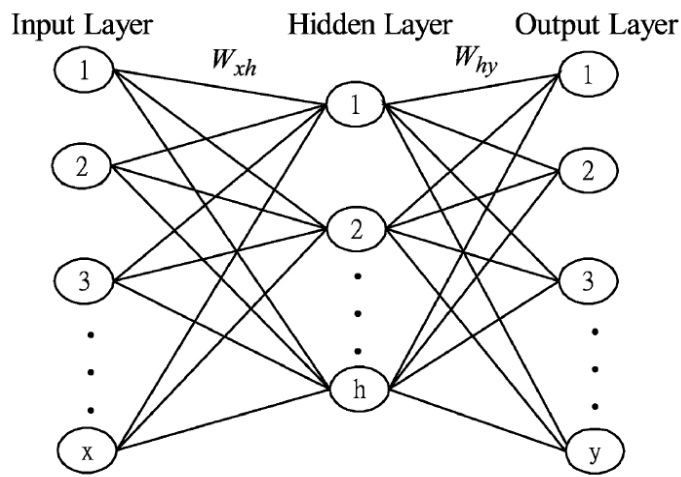

**Figure 8- Structure of a classic 3-layers artificial neural network**

Given a set of inputs neurons, $X_i (i = 1, 2, ..., x)$ the values are multiplied by the first set of interconnection weights, $(W_{xh})_{ih}$, where $(W_{xh})_{ih}$ is the connection weight from the $i$-th input neuron to the $h$-th hidden neuron. The summation of the products, $X_i (W_{xh})_{ih}$, can be written as (Rumelhart et al., 1986):

$$net_h = \sum_i X_i (W_{xh})_{ih} - \theta_{hh}, \qquad (LIII)$$

where $\theta_{hh}$ is the user-defined threshold of the $h$-th hidden neuron.

Each hidden neuron input is then transformed through the transfer function, such as the sigmoid function $f(x) = (1 + e^{-x})^{-1}$ to produce a hidden neuron output, $H_h$:

$$H_h = f(net_h) = \frac{1}{1 + e^{-net_h}}. \qquad (LIV)$$

Similarly, the output value between the hidden layer and the output layer are defined by:

$$\begin{cases} net_y = \sum_h H_h (W_{hy})_{hj} - \theta_{yj}, \\ \\ Y_j = f(net_j) = \frac{1}{1 + e^{-net_j}}, \end{cases} \qquad (LV)$$

where $(W_{hy})_{hj}$ is the connection weight from the $h$-th hidden neuron to the $j$-th output neuron, $\theta_{yj}$ the threshold of the $j$-th output neuron and $Y_j$ the value of output layer.

In general, the error at the output layer in the BPN model propagates backward to the input layer through the hidden layer in the network to obtain the final desired output. The gradient descent method is utilized to calculate the weights of the network and to adjust the weights of interconnections to minimize the output error.

The error function at the output neuron is defined as the least-square error function (Bishop, 1995):

$$E = \frac{1}{2} \cdot \sum_k (O_k - Y_k)^2, \qquad (LVI)$$

where $O_k$ and $Y_k$ are, respectively, the actual and predicted values of output neuron, and $k$ is the output neuron.

The gradient descent algorithm adapts the weights according to the gradient error, which is given by (Hirose, 2013):

$$\Delta W_{ij} = -\eta \times \frac{\partial E}{\partial W_{ij}}, \qquad (LVII)$$

where η is the learning rate and the general form of the $\partial E / \partial W_{ij}$ term is expressed by the following form (Bishop, 1995):

$$\frac{\partial E}{\partial W_{ij}} = -\delta_j^n \cdot Y_i^{n-1}. \qquad (LVIII)$$

Substituting (LVIII) into (LVII), we have the gradient error as:

$$\Delta W_{ij} = \eta \cdot \delta_j^n \cdot Y_i^{n-1}, \qquad (LIX)$$

where $Y_i^{n-1}$ is the predicted output value of sub-layer related to the connective weight ($W_{ij}$), and $\delta_j^n$ is the error signal, which is computed based on whether or not neuron $j$ is in the output layer. If neuron $j$ is one of the output neurons, then (Lee, 2004):

$$\delta_j = (O_j - Y_j) \cdot Y_j \cdot (1 - Y_j). \qquad (LX)$$

If neuron $j$ belongs to the hidden layer, then (Lee, 2004):

$$\delta_j = \left[ \sum_j \delta_j \cdot (W_{hy})_{hj} \right] \cdot H_h \cdot (1 - H_h), \qquad (LXI)$$

where $H_h$ is the value of the hidden layer.

Finally, the value of weight of the interconnected neuron can be expressed as follows (Lee, 2004):

$$W_{ij}^m = W_{ij}^{m-1} + \Delta W_{ij}^m = W_{ij}^{m-1} + \eta \cdot \delta_j^n \cdot Y_i^{n-1}. \qquad (LXII)$$

To accelerate the convergence of the error in the learning procedure, it is also possible to include in the equation an additional *momentum term* with a given momentum gain, α, as shown in the following equation (Jacobs, 1988):

$$W_{ij}^m = W_{ij}^{m-1} + \eta \cdot \delta_j^n \cdot Y_i^{n-1} + \alpha \cdot \Delta W_{ij}^{m-1}. \qquad (LXIII)$$

With this general model, artificial neural networks have been widely applied in tidal prediction and tsunami forecasting, where it is likely to have incomplete observed time series and efficient forecasting techniques, able to deal with missing data, like artificial networks, are required in order to obtain accurate predictions. Based on limited field data, the neural network method can predict hourly, daily, weekly or monthly tidal level accurately.

Neural networks were first applied to tide forecasting by using the field data of both diurnal and semidiurnal tides by Tsai and Lee (1999). However, their model was only applicable for the prediction of diurnal and semi-diurnal tides. In fact, mixed tides are more likely to occur in the field than diurnal and semi-diurnal tides (Lee et al., 1998). Lee et al. (2002) and Lee (2004) applied a neural network to predict different types of tides and found that the technique can be effective. However, their methods depend on harmonic parameters and cannot predict non-astronomical tidal level. Huang et al. (2003) developed a regional neural network water level model based on long-term tidal data from a NOAA station. Even though the non-astronomical components effect was significant, the model's prediction accuracy was good. Cox et al. (2002) and Lee (2006) addressed the prediction of tidal level including non-astronomical components. Lee and Jeng (2002) developed an advanced ANN model for tide forecasting using a short-term tidal record and considering all diurnal, semi-diurnal and mixed tides in their model. The data from three harbours in Taiwan were used as case studies, and the effects of the neural network structure, including training techniques and learning algorithms, were discussed in detail. Other recent works on the subject include (Liang et al., 2008; Deo, 2010; Wang et al., 2012).

Beltrami (2008; 2011) implemented algorithms based on the use of artificial neural networks in the software of bottom pressure recorders (BPRs) for the automatic, real-time detection of a tsunami within recorded signals. He compared his algorithms for tsunami detection to the one developed under the Deep-ocean Assessment and Reporting of Tsunamis (DART) program run by Mofjeld (1997) for the U.S. National Oceanic and Atmospheric Administration (NOAA). His results showed a consistent improvement in detection performance by using the ANN approach. The ANN algorithm by Beltrami (2008; 2011) is maybe the most efficient technology to date for the automatic and real-time detection of a tsunami. Its model is shown and discussed in the following section.

## 6.1 ANN algorithm by Beltrami (2008; 2001) for tsunami detection

At present, the most popular algorithm expressly designed to detect a tsunami in real-time within bottom pressure recorders (BPRs) is the one developed by Mofjeld (1997) under the U.S. NOAA's Deep-ocean Assessment and Reporting of Tsunamis (DART) program (Milburn et al., 1996).

The algorithm by Mofjeld (1997) uses a cubic polynomial to predict disturbing wave patterns. The predictions $\hat{\zeta}$ are updated every sampling interval (i.e. every 15 seconds). The polynomial is fitted to $n$-minute averages $\bar{\zeta}$ (centered at the $n/2$ minute) of bottom pressure observations $\zeta$ collected over the preceding 3 hours (Figure 9), and can be expressed as (Beltrami, 2008):

$$\hat{\zeta}(t') = \sum_{i=0}^{3} w_i \bar{\zeta}(t'' - i\Delta t), \qquad (LXIV)$$

where $t' = t + \frac{0.25}{60} h$, $t'' = t - n/(2 \cdot 60) h$, and $\Delta t = 1\ h$ ($t$ is the actual time). The coefficients $w_i$ are calculated by applying the Newton's forward divided difference formula, using the preceding temporal parameters.

Mofjeld (1997) suggests setting the time interval $n$ at 10 min in order to regularize the polynomial fitting point pattern by filtering out background noise typically recorded by BPRs, and to minimize the extent to which an actually detected tsunami indirectly influences the filtered signal by affecting the observation averages (Mofjeld, 1997).

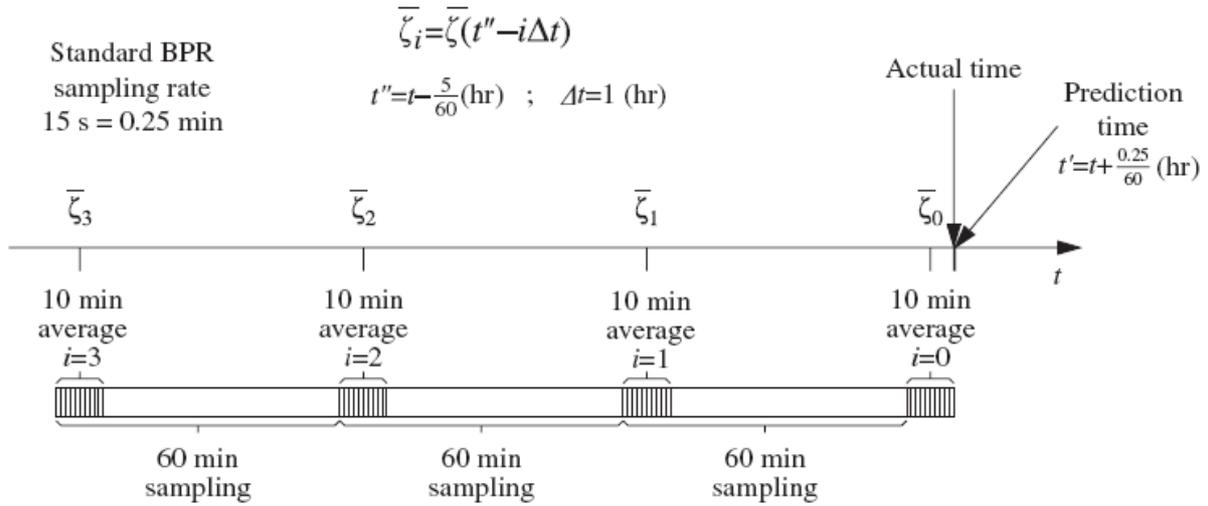

**Figure 9 - The DART tsunami-detection algorithm by Mofjeld (1997)**

The algorithm's prediction error depends both on the time interval *n* and on the magnitude of the disturbance to be filtered out. In the absence of background sea noise, this disturbance is mainly caused by the tide. Departure from a perfectly filtered signal (a zero signal) therefore depends on the measurement location. Whatever the time interval *n*, coefficients $w_i$ are calculated once and for all and a priori, by setting all the preceding temporal parameters; the implementation of Equation LXIV is therefore particularly simple. Furthermore, coefficients $w_i$ are calculated on the sole basis of temporal parameters. This makes the same set applicable to all BPRs, whatever their location.

The ANN tsunami detection algorithm by Beltrami (2008) emulates the main strengths of the DART algorithm by Mofjeld (1997) and is designed to improve the filtering performance methodology from a practical point of view. The ANN architecture and its characteristics are described in the following.

In order to update the prediction of disturbing fluctuations every 15 seconds, the ANN algorithm uses the feed-forward network shown in Figure 10. The two adaptive-weight layer ANN is characterized by 4 input units plus bias, 6 hidden units plus bias and one output unit (I4H6O1). The network's inputs consist of the same *n*-minute averages $\bar{\zeta}$ of bottom pressure observations $\zeta$ as those used from the DART algorithm by Mofjeld (1997). These values are pre-processed so as to re-scale them linearly in the range [0; 1]. More specifically, such a re-scaling is carried out by considering twice the original signal's maximum and minimum values.

The function represented by the network diagram (Figure 10) can be expressed as (Beltrami, 2008):

$$\hat{\zeta}(t') = \tilde{g}\left\{w_b^{(2)} + \sum_{j=1}^{6} w_j^{(2)} g\left[w_{bj}^{(1)} + \sum_{i=0}^{3} w_{ij}^{(1)} \bar{\zeta}(t'' - i\Delta t)\right]\right\}, \qquad (LXV)$$

where $t'$, $t''$, and $\Delta t$ are the same temporal parameters as those used in Eq. (LXIV), $w_{ij}^{(1)}$ and $w_{bj}^{(1)}$ are the adaptive weights connecting input units and bias to the hidden units, and $w_j^{(2)}$ and $w_b^{(2)}$ those connecting hidden units and bias to the output unit. Moreover, $g(\cdot)$ and $\tilde{g}(\cdot)$ represent the hidden and output unit activation functions. In particular, the logistic-sigmoid and the linear activation functions characterize the hidden units and the output unit, respectively. The structure of the described ANN algorithm is the experimental result of investigating the effect of different numbers of both hidden layers and units, and of different activation functions (see Beltrami (2008) for more details).

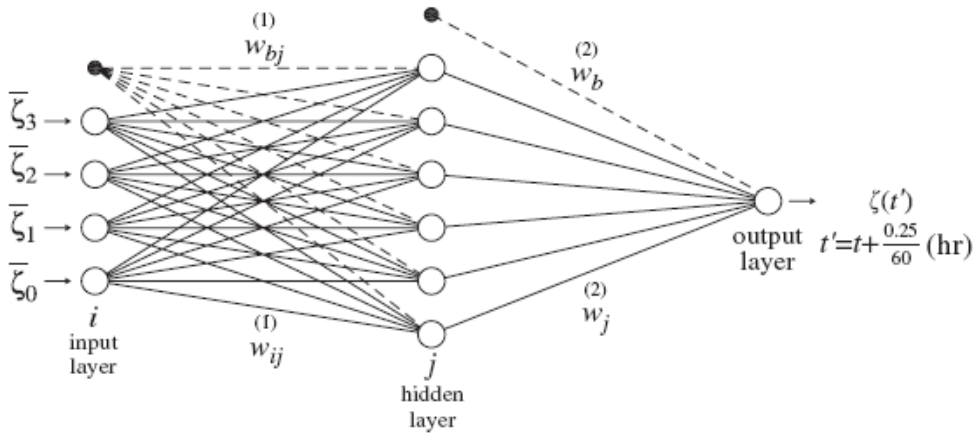

**Figure 10 - Diagram of the ANN algorithm by Beltrami (2008)**

Unlike the Mofjeld's coefficients of the cubic polynomial in Equation (LXIV), the adaptive ANN weights of Equation (LXV) result from the network's supervised learning. If a time series of actual observations $\zeta$ is available, it will be possible to present the network with an input array $[\bar{\zeta}]$ and a corresponding output vector $\{\zeta\}$, i.e. with a training set. The adaptive weights $[w]$ result from minimizing the error function chosen to express the difference between the calculated $\{\hat{\zeta}\}$ and the actually observed outputs $\{\zeta\}$. In particular, the ANN by Beltrami (2008; 2011) uses the mean square error as the error function (Bishop, 1995), and the standard back-propagation technique to calculate the error derivatives with respect to weights (Rumelhart et al., 1986; Hirose, 2013). In addition, the algorithm applies the Levenberg–Marquardt optimization scheme for iteratively adjusting the network weights at the end of each epoch by means of the calculated derivatives (Levenberg, 1944; Marquardt, 1963; Bishop, 1995).

The way in which the adaptive weights are calculated makes the network performance dependent neither on the time interval $n$ nor on the observed signal range. The interval $n$ could therefore be chosen for the purposes of minimizing the extent to which a detected tsunami influences the filtered signal indirectly. Actually, the efficiency of the supervised learning relies totally upon how accurately the training set represents all the possible disturbing fluctuations and their composition. If it succeeds in this, the ANN algorithm error is expected to be nearly zero. This is especially so in the case of disturbing fluctuations consisting of regular wavy patterns such as the tidal one (Beltrami, 2008).

On the other hand, the network training relies on actual data. This makes the resulting adaptive weights mainly tailored to a specific BPR location. In general, therefore, each BPR is characterized by a specific set of weights, although the same set might be used effectively in different BPRs if their locations are close enough to ensure that the devices are exposed to almost equal tidal and meteorological conditions (see Beltrami (2008; 2011) for more details).

Although it is possible to implement the network learning and executing modules in a BPR, the need for memory (to store all the data necessary for learning) and for power (to run the learning procedure and therefore update the adaptive weights regularly) makes this strategy inconvenient. A far more practical solution would be simply to implement Equation (LXV), i.e. just the executing module. Regularly downloading observed data makes it possible to carry out the network's supervised learning "off-BPR".

Furthermore, the training set can be updated on the basis of progressively longer observation time series. The resulting adaptive-weight array $[w]$ can be uploaded at the end of each new learning phase. Such a strategy clearly makes implementation simpler than that of the DART algorithm by Mofjeld (1997). It is worth noting that the length of the training set is not an absolute quality parameter of the supervised learning result (i.e. the adaptive-weight set). It is its accuracy in representing all the possible sea-level fluctuations (and their composition) that is of value.

The length of the observation time series, on the other hand, is important. The longer the available time series, the more informative the training set obtained by selecting representative tracks of observations will be and, therefore, the result of the learning phase as well. Although a time series extending over a few days is sufficient to get the astronomic tide feature, only a longer time series is likely to contain a statistically significant number of examples of the way in which stochastic fluctuations (e.g. those due to atmospheric-pressure field variations) superimpose on tidal and other regular long-period oscillations (e.g. planetary waves and gravitational normal modes on a geophysical-basin scale). More quantitatively, a training set consisting of a 15-day continuous time series (i.e. half a lunar cycle) is sufficient to capture tide behavior with extremely high precision and can therefore guarantee that the algorithm is fully and effectively operative. Selecting a 30-day continuous time series (i.e. a full lunar cycle) so as to include significant sea-level variations caused by meteorological perturbations clearly improves its effectiveness, however.

The experimental results reported in (Beltrami, 2008; 2011) show that the ANN algorithm's closer prediction of both tide and other regular patterns makes it capable of a better filtering performance than the DART algorithm by Mofjeld (1997) in most cases. Such an improvement may be more or less significant, depending on the range of the tide and the characteristics of the background sea noise. In particular, the higher the tidal range at the location of interest (and the lower the background sea noise), the greater the improvement in filtering performance by the ANN will be and, therefore, the feasible reduction of the detection threshold, as well. This further demonstrates that the ANN is a very effective automatic, real-time tsunami detection algorithm of ease implementation.

# 7   Conclusions and future research

In search for new directions for research in crisis and disaster management, the development of rapid tsunami detection algorithms and alerting systems shows tremendous potential. Tsunamis have hit, and will continue hitting our communities, businesses, and economies in oceanic and coastal areas. It is in everyone's interest to understand how it is possible to predict and manage them effectively and efficiently. A better detection of tsunamis will improve readiness, increase response speed and ease recovery, and primarily will save lives. The most important requisite of a tsunami detection algorithm consists of accurate analysis and forecasting of sea-level tide.

This paper has surveyed the state-of-art literature in tidal analysis and forecasting methods for tsunami detection to identify publication trends, issues worthy of further investigation, and aspects that have not yet been completely exploited.

The first conventional tidal forecasting methods, based on harmonic analysis, which uses the least squares method to determine harmonic parameters, have been discussed. In general, classic harmonic analysis methods require a large number of parameters and long-term measured data for precise tidal level predictions. Furthermore, depending on models based on the analysis of astronomical components, they can be inadequate when the contribution of non-astronomical components, such as the weather, is significant.

Other alternative approaches developed in the literature have been described in detail. They are able to deal with critical situations of the observed sea-level time series and provide predictions with the desired accuracy, with respect also to the length of the available tidal record. These methods include standard high or band pass filtering techniques, wavelets transform analysis methods, and artificial neural networks.

This paper is intended to provide the communities of both researchers and practitioners with a broadly applicable, up to date coverage of tidal analysis and forecasting methodologies that have proven to be successful in a variety of circumstances, and that hold particular promise for success for the development or the improvement of Tsunami Alerting Devices for the detection of tsunamis events. Appendix provides a starting point for crisis management researchers who would like to get a head start in understanding tsunami detection algorithms and alerting systems and meet others working in this area.

**Appendix I: useful resources**

- DART (Deep-ocean Assessment and Reporting of Tsunamis) website: http://nctr.pmel.noaa.gov/Dart/

- European Sea Level Service Monitoring website: http://www.eseas.org/

- Website of the National Oceanography Centre (NOC) of the Natural Environment Research Council: http://www.pol.ac.uk/

- Natural Environment Research Council (NERC) of United Kingdom: http://www.nerc.ac.uk/

- NOAA Center for Operational Oceanographic Products and Services: http://co-ops.nos.noaa.gov/index.shtml

- NOAA Center for Tsunami Research: http://nctr.pmel.noaa.gov/index.html

- Système d'Observation du Niveau des Eaux Littorales (SONEL): http://www.sonel.org/

- The British Oceanography Data Centre website: http://www.bodc.ac.uk/

- The Institute of Marine Science of the National Research Council (ISMAR-CNR): http://www.ismar.cnr.it/

- United States National Oceanic and Atmospheric Administration (NOAA): http://www.noaa.gov/

- Virginia Institute of Marine Science website: www.vims.edu

# Appendix II: abbreviations

- ANN: Artificial Neural Network

- AR: Auto-Regressive

- ARMA: Auto-Regressive Moving-Average

- BPR: Bottom Pressure Recorder

- CWT: Continuous Wavelet Transform

- DART: Deep-ocean Assessment and Reporting of Tsunamis

- DWDM: Dopplertyped Wave Directional Meter

- FIR: Finite Impulse Response

- IIR: Infinite Impulse Response

- MA: Moving-Average

- MSL: Mean Sea Level

- NERC: Natural Environment Research Council

- NOAA: U.S. National Oceanic and Atmospheric Administration

- NOC: National Oceanography Centre

- NOWPHAS: Japanese Nationwide Ocean Wave information network for Ports and HArbourS

- PARI: Port and Airport Research Institute of Japan

- PMEL: Pacific Marine Environmental Laboratory

- TEWS: Tsunami Early Warning Systems

- TGP: Tide Generating Potential